\documentclass[12pt]{article}
\usepackage[utf8]{inputenc}
\usepackage[T1]{fontenc}
\usepackage[english]{babel}
\usepackage{graphicx}
\usepackage{geometry}
\usepackage{bbold}
\usepackage{cancel}
\usepackage{amsmath}
\usepackage{amssymb}
\usepackage{xcolor}
\usepackage{cancel}
\usepackage{framed}
\usepackage[all]{xy}
\date{\today}

\begin{document}



\pagestyle{empty}
\rightline{LTH 1362}
\vskip 1.5 true cm  
\begin{center}  
{\large T-duality across non-extremal horizons}\\[.5em]
\vskip 1.0 true cm   
M.~M\'edevielle$^{1}$ and T.~Mohaupt$^{2}$\\[3ex] 
$^1${
Graduate School of Arts and Sciences, \\The University of Tokyo, \\Tokyo 153-8902, Japan\\
mmedevielle@g.ecc.u-tokyo.ac.jp
} \\[1em]
$^2${Department of Mathematical Sciences\\ 
University of Liverpool\\
Peach Street \\
Liverpool L69 7ZL, UK\\ 
Thomas.Mohaupt@liv.ac.uk \\[1em]
}
January 2, 2024. 
Revised: July 26, 2024
\end{center}
\vskip 1.0 true cm  
\baselineskip=18pt  
\begin{abstract}  
\noindent  
When applying T-duality to a generic, non-extreme Killing horizon,  T-duality is spacelike on one side and timelike on the other. We show, using simple examples from four-dimensional Einstein-Maxwell theory, that the image of the horizon is a singularity which can be understood as an interface between two different T-dual theories and their solutions. Using an embedding into type-II string theory, we show that the singularity occurs when scalars reach
the boundary of moduli space, resulting in a breakdown of the effective field theory due to the presence of tensionless strings.
\end{abstract}


\newpage
 \pagestyle{plain}
\tableofcontents

\section{Introduction}

T-duality is one of the features which makes string theory distinct. It states that
a perturbative string theory $A$, when compactified on a circle of radius $R$, is
equivalent to a perturbative string theory $B$, compactified on a circle of
radius $1/R$, where length is measured in string units $\sqrt{\alpha'}$. While the bosonic string is `T-selfdual,'
$A=B$,
we will be interested in type-II superstrings, where type-IIA and type-IIB are
T-dual to each other \cite{Dai:1989ua}, \cite{Dine:1988nrl}. The ten-dimensional type-II theories can be seen as two decompactification
limits, $R\rightarrow \infty$ and $R\rightarrow 0$, of a one-parameter family of type-II theories on $\mathbb{R}^{1,8} \times S^1_R$.
While the ten-dimensional theories are not equivalent, one still refers to them as `being T-dual' to each other. In this case T-duality is not an equivalence between theories, but a solution-generating technique, relating, for example, D-$p$-branes to D-$(p\pm 1)$-branes. 
By taking several directions to 
be compact one generates a larger discrete T-duality group.
We refer to \cite{Giveon:1994fu} for a review of T-duality. 

T-duality can
be formulated at the level of the effective (super-)gravity theories of the
massless modes. Here it can be derived by compactifying a theory
and decompactifying it in an alternative way (`taking $R\rightarrow 0$ in string
units'). As discussed above, when relating decompactified theories in this way, one does not
necessarily have an equivalence of theories, but a powerful solution generating
technique: by application of the `Buscher rules' to a solution of theory A
we obtain a solution of theory B.

T-duality shows that the framework of (pseudo-)Riemannian geometry used in 
general relativity is too narrow to provide a satisfactory geometric framework for 
string theory. It  suggests that $\sqrt{\alpha'}$ is a minimal length scale if we probe spacetime with strings. It also mixes 
the metric $g_{(10)}$ with the Kalb-Ramond field $b_{(10)}$ and the dilaton
$\phi_{(10)}$. This has motivated various approaches to define generalized
`stringy' geometries, where these data, and possibly also other massless 
string excitations, are treated on the same footing: generalized geometry \cite{Gualtieri:2003dx}, \cite{Hitchin:2010qz}, 
double field theory \cite{Hull:2006va}, \cite{Hull:2009mi}, as well as their `exceptional' and `heterotic' extensions which include fields outside the universal (NS-NS) sector \cite{Koepsell:2000xg}, \cite{Hull:2007zu}, 
\cite{PiresPacheco:2008qik}, 
\cite{Ashmore:2019rkx}. 
T-duality has
been used to construct various classes of string backgrounds which go outside the 
remit of Riemannian geometry \cite{Hull:2004in},\cite{Hull:2006va}.

In this paper we investigate an issue which arises when applying T-duality
to a very important class of Riemannian spacetimes, namely those which contain Killing horizons. While it was already observed in 
\cite{Rocek:1991ps} that T-duality maps horizons to singularities, not much further work seems
to have been done since then to follow up on this observation.\footnote{For certain two-dimensional backgrounds T-duality exchanges 
singularities and horizons, see Section 4 of \cite{Giveon:1994fu}.}
We will study generic, non-extremal Killing horizons, where
the square-norm of the Killing vector field 
has a simple zero at the horizon and thus changes between being timelike and being spacelike. From the viewpoint of a $(d-1)$-dimensional effective theory, obtained by dimensional reduction with respect to the Killing vector field, 
a non-extremal Killing horizon corresponds to an interface where spacetime signature changes:
for timelike reduction the lower-dimensioal theory has Euclidean signature, while for spacelike reduction it has Lorentzian signature. The T-dual of the $d$-dimensional solution is obtained by identifying and applying alternative dimensional lifts of the $(d-1)$-dimensional solutions. The resulting $d$-dimensional configuration solves the 
field equations of the T-dual theory, which in general is different from the original $d$-dimensional theory. 
Moreover, the T-dual theory is in general different for spacelike and timelike T-duality, so that in the T-dual
solution the horizon is mapped to an interface between solutions of two different theories. While all these operations
can be carried out for any gravitational theory coupled to matter, an embedding into string theory provides 
one with additional insights, as we will see in detail later. 
Note that, while we will thus provide string embeddings, we use
T-duality strictly as a solution generating technique for classical solutions of
field theories and of string theories. If one was to attempt to promote this to an
equivalence, it would require to take the isometric direction used for T-duality
to be compact. This in turn would require to include the dynamics of momentum 
and winding modes, and, in the case of timelike T-duality, to allow closed timelike
curves to be present. While we will make some speculative remarks at the end, 
we will otherwise not address such questions in this paper, and restrict ourselves 
to use T-duality to generate new solutions, and to investigate what happens to horizons under this solution-generating map.

Timelike T-duality was introduced in \cite{Hull:1998vg} where it was shown to map
the type-IIA/IIB string theories to two new superstring theories called type-IIB$^*$
and type-IIA$^*$. The partition functions of type-II$^*$ theories differ from those of the  
conventional theories by certain phases. At the level of the effective 
supergravity theories, the signs of the kinetic terms of all R-R fields
are reversed, while D-branes are replaced by E-branes which satisfy 
Dirichlet boundary conditions in real time. While this is highly unusual,
as long as the T-duality circle is kept finite, type-II$^*$ theories are
equivalent to type-II theories and all apparent pathologies are taken care
of by stringy gauge symmetries. When assuming that the decompactified
type-II$^*$ are well defined limits, and when adding S-duality into the mix,
one can generate a web of type-II string theories that extends over all
ten-dimensional space-time signatures \cite{Hull:1998ym}. Further features and potential 
pathologies of exotic type-II theories, as well as ways 
to cure them have been discussed in \cite{Dijkgraaf:2016lym} and \cite{Blumenhagen:2020xpq}. 
Lower-dimensional supergravity theories in non-Lorentizan signature, which arise from 
compactification of exotic string theories, as well as their solutions
have been studied recently in \cite{Gall:2018ogw},\cite{Cortes:2019mfa},\cite{Sabra:2021ugi},\cite{Sabra:2021omz},\cite{Sabra:2022xmf},\cite{Farotti:2023czm}.
As we will discuss later, one application of our work is to help clarify the physical validity
of dynamical transitions between conventional and exotic string theories. 

We will carry out our analysis from a four-dimensional perspective, and the reduction-lifting procedure we apply makes sense in the context of Einstein-Maxwell theory coupled to scalar fields. However, to properly interprete this procedure as T-duality, we need an embedding into string theory, which identifies one of the scalars as the four-dimensional dilaton and another as the universal string axion, dual to the four-dimensional Kalb-Ramond field. By performing the reduction-lifting procedure in the string conformal frame, we obtain four-dimensional Buscher rules which take the same form as in ten dimensions. In particular, T-duality corresponds to inverting the radius of a compactification circle  in string units, or, when decompactifying, inverting of the corresponding metric coefficient. 

We will use type-II and type-II$^*$ theories to provide a string embedding, and pick a particular solution of Einstein-Maxwell theory, the planar version of the non-extreme Reissner-Nordstrom solution, for concreteness. 
The geometric and thermodynamic properties of this solution, as well as its lift to ten-dimensional string theories and
eleven-dimensional M-theory have been studied in detail in \cite{Gutowski:2019iyo}, \cite{Gutowski:2020fzb}
and we will draw on this material where needed.
It will be obvious that the behaviour that we find is generic for spacetimes containing a non-extremal horizon. 
The planar Reissner-Nordstrom solution  describes 
a bouncing cosmology interpolating between a contracting and an expanding Kasner solution. The bounce is caused by the presence of timelike singularities, which reside in static regions, meeting the expanding and contracting regions along Killing horizons. 
We pick one static and one non-static region and apply timelike and spacelike T-duality, respectively. The T-dual theories are Einstein-dilaton-axion theories, which differ by a sign-flip in the axion kinetic term. To provide an embedding into string theory, we first observe that the Einstein-Maxwell and Einstein-dilaton-axion system both sit inside $\mathcal{N}=2$ supergravity coupled to a single hypermultiplet, with target space 
$U(1,2)/U(1) \times U(2)$ or 
$U(1,2)/U(1)\times U(1,1)$ depending on the sign of the axion kinetic term. These models are in turn what we may call the universal sector of type-II and type-II$^*$ superstring theory compactified on a Calabi-Yau threefold \cite{Medevielle:2021wyx}. They are universal in the sense that they are what remains if we consistently truncate
out all supermultiplets which contain moduli of the Calabi-Yau manifold. 
The scalars remaining in the `universal hypermultiplet' (UHM) are the four-dimensional dilaton $\varphi$, the universal axion $\tilde{\varphi}$, and two R-R scalars $\zeta, \tilde{\zeta}$. The embedding also shows that the Maxwell field is the universal
vector field present in any type-II/type-II$^*$ compactification, namely the one residing in the $\mathcal{N}=2$ supergravity multiplet. 

We will refer to the combined system 
of Einstein-Maxwell theory and the four scalars of the UHM with target $U(1,2)/U(1) \times U(2)$
as the EM-UHM system. We will show that this system is self-dual under spacelike T-duality (with certain subsystems mapped to each other), while under timelike T-duality it gets mapped to a system where the signs of the Maxwell term, and as well those of the kinetic terms of the R-R scalar $\zeta, \tilde{\zeta}$ have been flipped. This changes the scalar target geometry to $U(1,2)/U(1)\times U(1,1)$. We will refer to this theory as the twisted
EM-UHM system, and use the notation 
(EM-UHM)$_{-}$ to indicate the sign flips. 
The twisted EM-UHM system is the bosonic part of a twisted $\mathcal{N}=2$ supergravity theory, which realises a twisted version of the standard four-dimensional $\mathcal{N}=2$ supersymmetry algebra with R-symmetry group $U(1,1)$ \cite{Cortes:2015wca}. In this theory, the target space geometry of hypermultiplets is para-quaternionic K\"ahler instead of quaternionic K\"ahler. The twisted EM-UHM system is the universal sector of type-II$^*$ compactifications on Calabi-Yau threefolds \cite{Medevielle:2021wyx}. It is self-dual under spacelike T-duality and gets mapped to the untwisted version under timelike T-duality. 
By performing spacelike and timelike reductions and liftings for both models, we will obtain the explicit Buscher rules relating the  fields. 

\begin{figure}
\xymatrix{
\mathcal{N}=0 & \mathcal{N}=2 & \mathcal{N}=8& D=10 \\
\mbox{EM} \ar[dr] \ar@{<->}[dd]^{T} &  &        & \mbox{type IIA} \ar@{<->}[dd]^T \\
      &  \mbox{EM}+\mbox{UHM}  \ar@/^1pc/[rru]^{CY_3} \ar[r] \ar@/_1pc/[rrd]_{CY_3}&
       \mbox{$\mathcal{N}=8$\,SG}  \ar[ur]_{\mathrm{Torus}} \ar[dr]^{\mathrm{Torus}} &   \\ 
\mbox{E-Dil-Ax} \ar[ur]&  &  &  \mbox{type IIB} \\
}
\caption{Action of spacelike T-duality on various models discussed in the text. Double arrows with a `T' denote relations  by spatial T-duality, the other arrays are 
embeddings and dimensional liftings. Absence
of a T-duality arrow indicates that the model
is self-dual under spatial T-duality. A similar diagram (not shown) relates twisted 
versions of the various models, which differ 
from the models shown here by sign flips for the kinetic terms of some of the fields. \label{fig:1}}
\end{figure}
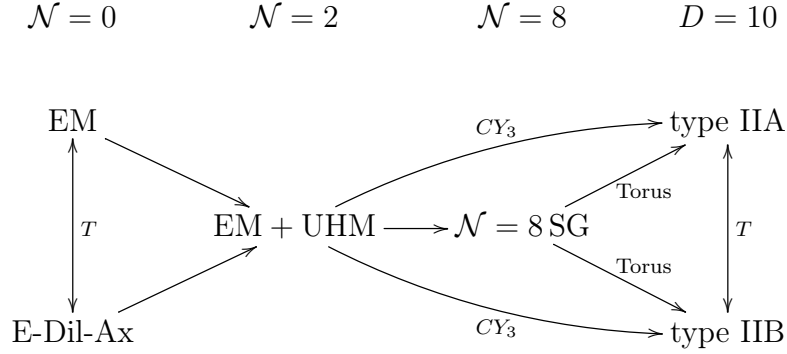

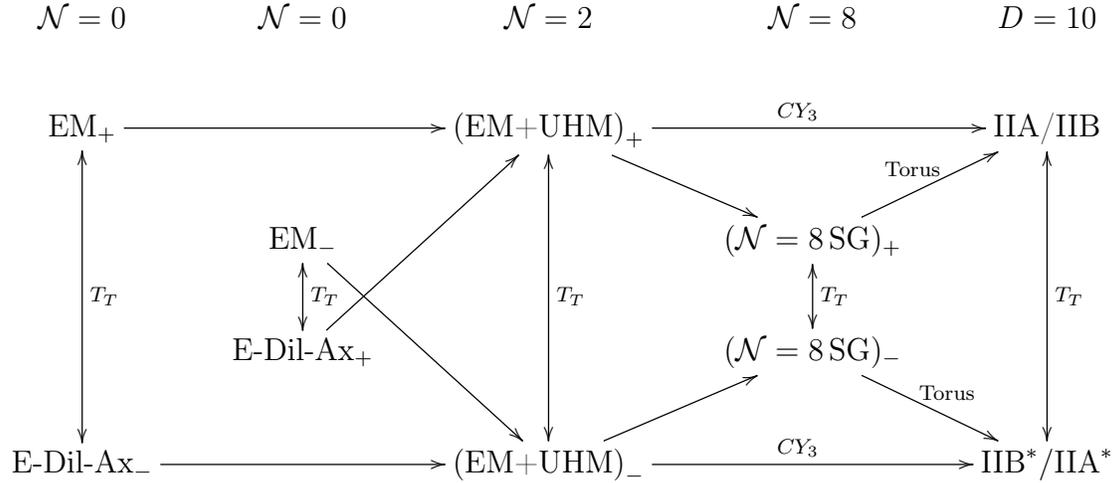
\begin{figure}
\xymatrix{
\mathcal{N}=0 &\mathcal{N}=0  & \mathcal{N}=2 & \mathcal{N}=8 & D=10 \\
\mbox{EM}_+  \ar@{<->}[ddd]^{T_T} \ar[rr] & &\mbox{(EM+UHM)}_+  \ar@{<->}[ddd]^{T_T}  \ar[rr]^{CY_3} \ar[rd] & 
 & \mbox{IIA/IIB} \ar@{<->}[ddd]^{T_T}\\
 & \mbox{EM}_-  \ar@{<->}[d]^{T_T}  \ar[rdd]  &  & (\mbox{$\mathcal{N}=8$\,SG})_+ \ar@{<->}[d]^{T_T}  \ar[ru]^{\mathrm{Torus}} &  \\
& \mbox{E-Dil-Ax}_+  \ar[ruu] &  & (\mbox{$\mathcal{N}=8$\,SG})_-  \ar[rd]^{\mathrm{Torus}} & \\
\mbox{E-Dil-Ax}_-  \ar[rr] &  & \mbox{(EM+UHM)}_- \ar[rr]^{CY_3} \ar[ru] &  &  \mbox{IIB}^*/\mbox{IIA}^*\\
}
\caption{Action of timelike T-duality on various models
discussed in the text. Arrows with a $T_T$ indicate the action of timelike T-duality, the other arrows indicate embeddings and dimensional lifts. Subscripts $\pm$ distinguish between the standard and twisted versions of the models. \label{fig:2}
}
\end{figure}

We remark that instead of using Calabi-Yau compactifications, we can also embed our models using toroidal compactifications. A toroidal compactification
of type-II string theory leads to $\mathcal{N}=8$ supegravity, which can be truncated to $\mathcal{N}=2$ supergravity. The toroidal reduction of
the type-IIA/IIB theories leads to the standard $\mathcal{N}=8$ supergravity theory with scalar target space $E_{7(7)}/SU(8)$, while toroidal reduction of the type-IIA$^*$/IIB$^*$ theories leads to a twisted version with scalar target space $E_{7(7)}/SU(4,4)$ \cite{Hull:1998vg}. 
A toroidal embedding has the advantage that one can explicitly lift four-dimensional solutions to ten or eleven dimensions, and identify the corresponding brane configurations. The planar Reissner-Nordstrom 
solution lifts to the same brane configuration that underlies the `STU-black hole', namely a D4-D4-D4-D0 system in type-IIA 
and D5-D1 system augmented with a pp-wave and
a Taub-NUT soution in type-IIB. The relations between
the various models discussed above, as well as the action of spacelike and timelike T-duality on them, is summarized in Figure \ref{fig:1} and \ref{fig:2}.

We will see that the T-dual solutions become singular at both the positions of 
the horizon and of the singularity of the original solution.
While from the purely gravitational point of view, these are just naked singularities,
a string theory embedding will allow us to make sense out of them:
they are related to the run-away behaviour of scalar fields, which suggests an interpretation 
in terms of the swampland distance conjecture \cite{Ooguri:2006in}. This conjecture states that whenever scalar
fields (`moduli') in a string-effective field theory approach the boundary of their target space, there
should be an infinite tower of states in the full string theory which becomes massless in this limit,
resulting in a breakdown of the effective field theory. By choosing a suitable string embedding and
duality frame, we will argue that all singularities observed in the T-dual solution are related
to limits where such a tower arises through strings becoming tensionless. While the singularities 
appear at infinite distance in moduli space, they are at finite distance in spacetime, which 
raises the question of what happens at these points. We will make some concrete proposals how
these questions can be addressed based on the results of our work in the conclusions section.

This paper is organised as follows. In Section 2 we provide the necessary background on
T-duality and its relation to the supergravity c-map. In Section 3 we derive the 
four-dimensional Buscher rules for our EM-UHM model using the reduction/lifting method.
In Section 4 we apply these results to T-dualize the 
planar cosmological solution and to investigate its properties in detail. Using an embedding of the Einstein-Maxwell solution into type-IIA string theory, its T-dual image embeds into type-IIB$^*$ and type-IIB depending on which side of the horizon we are. We use these embeddings to argue that the singularities of the T-dual solution are related to tensionless strings. In Section 5 we apply the same procedure to the planar black hole solution of a twisted Einstein-Maxwell theory where the sign of the Maxwell field has been flipped. This model can be embedded into type-IIA$^*$ and we briefly discuss its relation to the planar cosmological solution, when embedded into type-IIA. We conclude with an outlook on further research. Some details have been relegated to appendices.

\section{Background}

\subsection{Solution-generation through dimensional 
reduction/lifting}

This paper can be read from two perspectives. The first is
the one of classical gravity, where we 
provide and apply a solution-generating 
technique for four-dimensional Einstein-Maxwell 
theory coupled to a non-linear sigma model with four real scalar 
fields $\varphi, \tilde{\varphi}, \zeta, \tilde{\zeta}$ parametrising the symmetric space $U(2,1)/(U(2)\times U(1))$. The corresponding action is
\begin{align}
    \mathcal{S}_4&=\int d^4x \sqrt{\hat{g}_E} 
    \left( \hat{R}_E-\frac{1}{2}\hat{g}_E^{\mu\rho}\hat{g}_E^{\nu\lambda}\hat{F}_{\mu\nu}\hat{F}_{\rho\lambda}-2\hat{g}_E^{\mu\nu}\partial_{\mu}\varphi\partial_{\nu}\varphi  \right.
    \nonumber \\
    &-2e^{4\varphi}
    \left[\right. \partial^{\mu}\tilde{\varphi}+\frac{1}{2}(\zeta\partial^{\mu}\tilde{\zeta}-\tilde{\zeta}\partial^{\mu}\zeta)]\left[\right.
    \partial_{\mu}\tilde{\varphi}+\frac{1}{2}(\zeta\partial_{\mu}\tilde{\zeta}-\tilde{\zeta}\partial_{\mu}\zeta)] \nonumber \\
    &\left. -e^{2\varphi}\left[\partial_{\mu}\zeta \partial^{\mu}\zeta+ \partial^{\rho}\tilde{\zeta}\partial_{\rho}\tilde{\zeta}\right] \right)
    \label{action}
\end{align}
It is a special case of the class of actions considered in \cite{Medevielle:2021wyx}. Apart from truncating out all but one vector field and all but four scalars, we have rescaled the field $\varphi$ (which from the string perspective considered
below is the four-dimensional dilaton) by $\varphi \rightarrow -2\varphi$ and there is an overall factor of 2. Anticipating that we will apply dimensional reduction, in this section the four-dimensional metric and vector field 
carry a `hat', and the letter `E' denotes that we are in the Einstein
conformal frame. 

The class of solutions we start with has trivial scalars and thus
is a solution to Einstein-Maxwell theory. It takes the form 
\begin{equation}
\label{solution}
ds^2 = - f(r) dt^2 + \frac{1}{f(r)}dr^2 + r^2 (dx^2 + dy^2) \;,\;\;F = - \frac{e}{r^2} dt \wedge dr \;,
\end{equation}
where 
\begin{equation}
f(r) = - \frac{2M}{r} + \frac{e^2}{r^2} \;,
\end{equation}
with a mass-like parameter $M>0$ and electric
charge $e$, such that $M^2>e^2$. For later reference we note that a gauge potential 
is given by $A=-\frac{e}{r}dt$.
The solution has planar symmetry in the two spacelike coordinates $x,y$. For $0<r<r_* = \frac{e^2}{2M}$ the coordinate $r$ is spacelike, while $t$ is timelike and the solution is static with timelike Killing vector field $\xi=\partial_t$. For $r_*<r< \infty$,
$r$ is a timelike coordinate and the isometry $\partial_t$ is spacelike. Thus $r=r_*$
is a Killing horizon of the non-extremal type, where square-norm of the Killing vector field has a simple zero and changes between spacelike and 
timelike. For $r\rightarrow \infty$ the solution is asymptotic to a Kasner cosmological solution, while $r=0$ is a curvature singularity. 
This cosmological solution is the planar symmetric cousin of the familiar non-extremal Reissner-Nordstr\"om solution. Its global geometric structure
and the thermodynamics of its horizon are described 
in detail in \cite{Gutowski:2019iyo,Gutowski:2020fzb}, where
this solution appeared as a special case of a class of solutions
for the so-called STU-model, an $\mathcal{N}=2$ supergravity theory with three vector supermultiplets. Within this class of models, solutions to Einstein-Maxwell
theory can be obtained by setting its three complex scalars constant
and by setting its four vector fields proportional to each 
other \cite{Gutowski:2019iyo}, \cite{Gutowski:2020fzb}.

Compared to the spherically symmetric Reissner-Nordstr\"om solution the planar version lacks the 
exterior, asymptotically flat region, so that the 
intermediate dynamical region between the inner and outer horizon  has now become the exterior region, making this a cosmological rather than a black hole 
solution. In the extremal limit of the planar solution
the horizon and the exterior region are pushed off to infinity, leaving a static solution with a naked singularity, 
see Figure 5 of \cite{Gutowski:2019iyo}. We only consider the non-extremal solution in the following, since we are interested in the effects in T-dualizing in the presence of a Killing horizon. 
By maximal analytical continuation one obtains two copies of the static interior and two copies of the dynamical exterior region. The resulting Penrose diagram is the one of the maximally extended Schwarzschild spacetime, rotated by 90 degrees, see Figure 3 of \cite{Gutowski:2019iyo}. After picking a global time orientation, it describes a bouncing cosmology, which is complete for timelike geodesics. The static regions, sourced by time-like curvature singularities (which can be related to certain branes upon embedding into string theory) connect an early contracting to a late expanding cosmology, making it a cosmological bounce which is timelike geodesically complete. 

To this solution we will apply a solution-generating technique which uses dimensional reduction with respect to the 
Killing vector field $\partial_t$, 
followed by an `alternative dimensional lift' or `oxidation', which 
exchanges the degrees of freedom of metric and Maxwell field with 
those of the four scalars. In the dynamical region, where the Killing 
vector field is spacelike, the reduced $(1+2)$-dimensional theory is 
the bosonic part of $\mathcal{N}=4$ supergravity coupled to two hypermultiplets.
The dynamical degrees of freedom are eight scalar fields, which parametrize the 
symmetric space 
\begin{equation}
\mathcal{N} =
\frac{U(2,1)}{U(2)\times U(1)} \times 
\frac{U(2,1)}{U(2)\times U(1)}\;.
\end{equation}
This follows immediately from the results of \cite{Cortes:2015wca}
and \cite{Medevielle:2021wyx}.
By inspection it is obvious that 
there exists an alternative lift to $1+3$ dimensions, with the roles 
of four-dimensional scalar and non-scalar fields exchanged. 
Since the Killing vector field changes type at the
horizon, the solution generating reduction in the static region is timelike, leading to an effective three-dimensional Euclidean theory. We will come back to this below.

\subsection{T-duality and the c-map}

The second perspective that we can take is the one of ten-dimensional type-II superstring 
theory, into which the solutions can be embedded. Specifically, an
explicit uplift was given in \cite{Gutowski:2019iyo} for the static 
part of the solution \eqref{solution}, with the type-IIB D5-D1-pp-wave system and the 
M-theory M5-M5-M5-pp-wave system among the possible lifts. 
These lifts use that the EM-UHM system (using terminology
and acronyms introduced in the introduction)
can be embedded into the maximal (ungauged) four-
dimensional $N=8$ supergravity theory, which in turn is the low-energy 
effective field theory of the massless modes for type-IIA/IIB string 
theory compactified on a six-torus $T^6$. In this context, the 
solution generating technique described above amounts to applying the 
T-duality between type-IIA string theory compactified on $T^6\times 
S^1_R$ and type-IIB string theory compactified on $T^6\times 
S^1_{1/R}$, where $R$ is the radius of an additional circle, measured 
in string length units, followed by taking the alternative 
compactification limit $R\rightarrow 0$ to obtain a duality between 
two effective four-dimensional theories and their solutions.

Apart from being a subsector of $\mathcal{N}=8$ supergravity, the EM-UHM system is the bosonic part 
of $\mathcal{N}=2$ supergravity coupled to a single hypermultiplet, and can be considered as the universal sector that any type-II compactification on a Calabi-Yau threefold contains. In general, compactification of type-IIA superstring theory on a Calabi-Yau threefold with Hodge numbers $(h_{1,1}, h_{2,1})$
results in $\mathcal{N}=2$ supergravity coupled to $n_V = h_{1,1}$ vector multiplets and $h_{2,1}+1$ hypermultiplets
\cite{Bodner:1990zm}. To leading order in $\alpha'$, one of the hypermultiplets parametrizes a totally geodesic submanifold of the hypermultiplet moduli space which is isomorphic to 
$U(2,1)/(U(2)\times U(1))$, so that a consistent truncation to this single hypermultiplet is possible.\footnote{See \cite{Aspinwall:2000fd} for subtleties arising through $\alpha'$-corrections.}
 This hypermultiplet is called the universal hypermultiplet (UHM) and contains the 
IIA dilaton $\varphi$, the universal axion $\tilde{\varphi}$ obtained by Hodge-dualizing the four-dimensional Kalb-Ramond $B$-field, and two scalars $\zeta, \tilde{\zeta}$ descending from the ten-dimensional 
R-R sector (Ramond-Ramond sector). These scalars are universal in 
the sense that even in the case of a rigid complex structure there are
at least two homological three-cycles associated with the holomorphic
top form and its complex conjugate.\footnote{In terms of Betti numbers
$b_i$ and Hodge numbers $h_{i,j}$: $b_3 = 2 + 2 h_{1,2}$ for Calabi-Yau
threefolds.} Reducing the R-R three-form potential $C_3$ over these
cycles gives rise to two real scalars. 
The scalars in the other (non-universal) hypermultiplets parametrize deformations of the complex structure of the Calabi-Yau manifold, together with further R-R scalars obtained by reducing 
$C_3$ on the other $2h_{2,1}$ homological three-cycles. Truncating these fields out thus amounts to freezing the complex structure and the non-universal R-R moduli.  The theory also contains $h_{1,1}+1 = n_V+1$ vector fields one of which, dubbed the graviphoton, belongs to the
$\mathcal{N}=2$ supergravity multiplet while the others sit in the 
$n_V= h_{1,1}$ vector multiplets. The (complex) scalar fields in IIA vector multiplets 
parametrize the (complexified) K\"ahler structures of the Calabi-Yau threefold. Freezing them out corresponds to fixing the K\"ahler structure as well as the moduli of the ten-dimensional Kalb-Ramond field. After truncating out the 
vector multiplets, a single vector field remains  which we 
refer to as the Maxwell field. We remark that while our model with $n_V=0, n_H=1$ formally corresponds to the case $h_{1,1}=h_{2,1}=0$, there is no such 
Calabi-Yau threefold, since one can always at least change the overall size and hence $h_{1,1} \geq 1$. However, the 
EM-UHM system is a consistent truncation where all
non-universal scalars have been frozen and only one vector field
has been kept. We have focused on the type-IIA case. If one compactifies type-IIB string theory on the same Calabi-Yau manifold
one obtains a different $\mathcal{N}=2$ supergravity theory which has 
$n'_V = h_{2,1}$ vector multiplets and $n'_H = h_{1,1}+1$ hypermultiplets. It is these two theories that are related
by T-duality over a circle transverse to the Calabi-Yau 
manifold \cite{Cecotti:1988qn}.\footnote{Note that this needs to be distinguished from mirror symmetry, which is an equivalence between type-IIA and type-IIB compactified on two different Calabi-Yau threefolds 
$X_6$, $\tilde{X}_6$ with
Hodge numbers related by $\tilde{h}_{1,1}= h_{2,1}$, $\tilde{h}_{2,1}=h_{1,1}$.}

In the context of four-dimensional $\mathcal{N}=2$ supergravity and of type-II Calabi-Yau compactifications, the action of T-duality on a circle transverse to the Calabi-Yau space acts on the scalar fields by the so-called c-map 
\cite{Cecotti:1988qn}, \cite{Ferrara:1989ik}. Given  type-IIA and type-IIB theories compactified on the same Calabi Yau manifold $X_6$, T-duality states that
type-IIA on $X_6\times S^1_R$ is equivalent to type-IIB on $X_6\times S^1_{1/R}$, where $R$ is the radius of a circle orthogonal to $X_6$
measured in string length units. By taking the decompactification 
limits $R,1/R \rightarrow \infty$, one obtains a map between
type-IIA and type-IIB compactified on $X_6$, which can be used to 
generate solutions of the type-IIB effective supergravity theory
from those of the type-IIA theory, and vice versa.
Starting from type-IIA with $n_V=h_{1,1}$ vector multiplets and $n_H=h_{2,1}+1$ hypermultiplets, spacelike dimensional reduction results in a theory with $(h_{1,1} + 1)+ (h_{2,1}+1)$ hypermultiplets: after reduction to three dimensions, the $h_{1,1}$ vector multiplets 
can be dualized into $h_{1,1}$ hypermultiplets, while the four local 
degrees of freedom of the four-dimensional Einstein-Maxwell system 
organise themselves into a hypermultiplet with scalar manifold
$U(2,1)/(U(2)\times U(1))$, which is a totally geodesic
subspace of a combined scalar manifold $N_{4h_{1,1}+4}$ hosting
all degrees of freedom descending from the four-dimensional 
supergravity and vector multiplets. This manifold is
quaternionic-K\"ahler \cite{Ferrara:1989ik}.
The four-dimensional hypermultiplet manifold
$N'_{4h_{2,1}+4}$ which also is a quaternionic K\"ahler manifold  reduces trivially, 
so that the resulting total scalar manifold of the reduced
three-dimensional theory 
is a metric product 
\[
\mathcal{N} = N_{4h_{1,1}+4} \times N'_{4h_{2,1}+4}
\]
of 
two quaternionic-K\"ahler manifolds, as required for consistent 
coupling to three-dimensional supergravity. 
This in turn allows an alternative 
dimensional uplift to a four-dimensional $\mathcal{N}=2$ supergravity theory
with $n'_V=h_{2,1}$ vector multiplets and $n'_H=h_{2,1}$ hypermultiplets, which is the low energy effective field theory
of type-IIB string theory compactified on $X_6$.

In this process the four-dimensional supergravity 
multiplet gets mapped to the universal hypermultiplet. Therefore the
EM-UHM model is the smallest theory
for which this duality makes sense, and, moreover, the 
EM-UHM model 
is self-dual since the total numbers $n_V=0, n_H=1$ of vector and 
hypermultiplets does not change and the two factors of three-dimensional
scalar manifold are isometric to each other.

\subsection{Timelike T-duality and the image of a Killing horizon}

Since the spacetime \eqref{solution} has a non-extremal Killing 
horizon at $r=r_*$, the region $0<r<r_*$ has to be treated separately.
The Killing vector field is now timelike, so that by reduction we obtain a three-dimensional Euclidean effective theory.
Timelike dimensional reduction is a standard technique for generating stationary solutions 
\cite{Galtsov:1998yu}, \cite{Stelle:1998xg}. The maximal Euclidean supergravity theories 
are known from dimensional reduction of eleven-dimensional supergravity and ten-dimensional 
type-IIB supergravity on Lorentzian tori \cite{Cremmer:1998em}, \cite{Hull:1998vg}.
Euclidean
$\mathcal{N}=2$ supergravity in four and three dimensions features a variant
of special geometry where complex structures are replaced by 
para-complex structures. These geometries were developed 
in depth in \cite{Cortes:2003zd}, \cite{Cortes:2005uq}, \cite{Cortes:2009cs},
\cite{Cortes:2015wca},see also \cite{LopesCardoso:2019mlj}
for a comprehensive review. This work has
been extended to all four-dimensional and three-dimensional
signatures in \cite{Medevielle:2021wyx}. Scalar geometries 
depend on the dimension and signature in a way that is encoded
in the corresponding supersymmetry algebra, 
specifically in its R-symmetry group \cite{Gall:2021tiu}. In the following
we will rely on this work for background and will sometimes quote
results from it. For the specific purpose of this paper the relevant 
point is the variation of various relative signs between terms in 
the Lagrangian. 

The special geometry of four-dimensional $N=2$ vector multiplets
coupled to supergravity is projective special K\"ahler, see
\cite{LopesCardoso:2019mlj} for a review which uses our conventions.
While spacelike reduction leads, after inclusion of
the reduced degrees of freedom of the supergravity multiplet,
to a quaternionic-K\"ahler manifold \cite{Ferrara:1989ik},
a timelike reduction leads to a so called para-quaternionic
K\"ahler manifold, where the signs of the kinetic terms of half of
the scalars have been flipped \cite{Cortes:2019mfa} . From the 
perspective of type-II Calabi-Yau compactifications, these
are precisely the R-R scalars. If one just reduces the 
four-dimensional supergravity multiplet over time, the
resulting manifold is one of the two Euclidean versions
of the UHM, namely $U(2,1)/(U(1,1)\times U(1))$.
Since the four-dimensional hypermultiplet reduces trivially, 
timelike reduction of the EM-UHM theory results in a three-dimensional 
Euclidean theory with scalar manifold
\begin{equation}
\label{N-prime}    
\mathcal{N}' = 
\frac{U(2,1)}{U(1,1)\times U(1)} \times \frac{U(2,1)}{U(2)
\times U(1)} \;.
\end{equation}
This follows immediately from the results of \cite{Cortes:2015wca}
and \cite{Medevielle:2021wyx}.
If one performs an alternative  dimensional uplift over time
it is the second factor which will give rise to the four-dimensional
graviton and Maxwell field, while the four-dimensional scalar fields
now parametrize the first factor. Thus the EM-UHM
model is not self-dual under timelike T-duality but is mapped to a dual 
theory where various signs have been flipped. According to 
\cite{Medevielle:2021wyx} the resulting four-dimensional
action is
\begin{align}
    \mathcal{S}_4&=\int d^4x\sqrt{\hat{g}_E}\Big(\hat{R}_E \color{red} {\bf +}\color{black} \frac{1}{2}\hat{g}_E^{\mu\rho}\hat{g}_E^{\nu\lambda}\hat{F}_{\mu\nu}\hat{F}_{\rho\lambda}-2\hat{g}_E^{\mu\nu}\partial_{\mu}\varphi\partial_{\nu}\varphi \nonumber \\
    &-2e^{4\varphi}\left[\right. \partial^{\mu}\tilde{\varphi}+\frac{1}{2}(\zeta\partial^{\mu}\tilde{\zeta}-\tilde{\zeta}\partial^{\mu}\zeta)]\left[\right.\partial_{\mu}\tilde{\varphi}+\frac{1}{2}(\zeta\partial_{\mu}\tilde{\zeta}-\tilde{\zeta}\partial_{\mu}\zeta)] \nonumber \\
    &\color{red}+\color{black} e^{2\varphi}\left[\partial_{\mu}\zeta\partial^{\mu}\zeta+\partial^{\rho}\tilde{\zeta}\partial_{\rho}\tilde{\zeta}\right]\Big). \label{action2}
\end{align}
Observe the sign flips, indicated in red, of the kinetic terms of the Maxwell field and of the scalars
$\zeta, \tilde{\zeta}$, compared to the EM-UHM theory  \eqref{action}.
We refer to \eqref{action2} as the twisted EM-UHM theory, denoted (EM-UHM)$_-$, see 
Figure $\ref{fig:2}$.
It comprises the universal sector (supergravity multiplet and
universal hypermultiplet) of type-II$^*$ superstring theories compactified on a Calabi-Yau threefold. 
Due to the sign flips 
the bosonic Lagrangian  \eqref{action2} completes into a Lagrangian invariant 
under a twisted version of the standard $\mathcal{N}=2$ supersymmetry algebra,
which has R-symmetry group $U(1,1)\simeq SU(1,1) \times U(1)$ rather
than $U(2)\simeq SU(2) \times U(1)$ \cite{Cortes:2019mfa}.
In this twisted Lorentz signature theory, vector multiplet 
geometry is still projective special K\"ahler, but the vector fields
have inverted kinetic terms, while the hypermultiplet geometry is
para-quaternionic K\"ahler. 
The twisted EM-UHM model \eqref{action2} is self-dual
under spacelike T-duality, since dimensional reduction over space 
leads to a three-dimensional theory with scalar target space
\begin{equation}
\mathcal{N}'' = \frac{U(2,1)}{U(1,1)\times U(1)} \times \frac{U(2,1)}{U(1,1)
\times U(1)} \;,
\end{equation}
as follows immediately from the results of \cite{Cortes:2015wca}
and \cite{Medevielle:2021wyx}.

\subsection{Summary of this section}

In this section we have reviewed the c-map, including its extension to 
timelike reductions and lifts. When applied to Einstein-Maxwell theory
coupled to the UHM, T-duality amounts to exchanging the roles of the 
two UHMs which carry the dynamical degrees of freedom after dimensional reduction. We have taken
the supergravity perspective, using the Einstein conformal frame.
In this parametrization of the four-dimensional actions, the assignments
of fields to supermultiplets is clear. However, while T-duality can 
be viewed as a reduction/lifting procedure applicable to any gravitational
theory, it fundamentally is a `stringy' duality, and these feature 
only become manifest when using the string conformal frame. We will
turn to this perspective in the next section.

\section{Four-dimensional Buscher rules}

We now turn to the derivation for the Buscher rules for the EM-UHM
model and its twisted version. 
In order to work out the T-dual of the solution \eqref{solution}, 
we could in principle work with the actions \eqref{action} and
\eqref{action2}. The $\mathcal{N}=2$ Lagrangians resulting from Calabi-Yau
compactifications of all type-II string theories, including those of
non-Lorentzian signature, can be found in \cite{Medevielle:2021wyx} together
with those of all possible spacelike and timelike reductions to three
dimensions. From this information one can build a dictionary between
the fields of any two four-dimensional theories that are related by 
a spacelike, timelike or mixed (signature changing) T-duality. By 
applying these rules to any solution of one theory, one obtains a corresponding
solution of the T-dual theory. 

However, the Lagrangians used in \cite{Medevielle:2021wyx} were all given 
in the Einstein frame, while T-duality looks more natural in the 
string frame. By committing to embedding the solution \eqref{solution} into a
type-II string theory (which we take to be IIA for concreteness), with $\varphi$ as four-dimensional dilaton 
and $\tilde{\varphi}$ as the universal axion, we know how to rewrite
the four-dimensional effective 
theory in terms of the four-dimensional string frame. Then, by going
through the reduction/lifting process with the dual theory again 
be written in terms of the four-dimensional string frame, we
expect to obtain four-dimensional Buscher rules which resemble
the ten-dimensional ones. In particular, since T-duality corresponds
to inversion of the radius of a compactified isometric direction `*', 
we expect that the corresponding four-dimensional string frame 
metric coefficient transforms as $g_{**} \mapsto g_{**}^{-1}$. 
While the bosonic string frame NS-NS action takes the same form in 
any dimension, the R-R sector is non-universal, and therefore we have
to go through the reduction/lifting procedure to obtain the full set 
of Buscher rules.

\subsection{Spacelike reduction of the Einstein-Maxwell-UHM sytem}

In this section we derive the four-dimensional Buscher rules
for T-dualizing the EM-UHM system over space. Before reducing to three
dimensions, we transform 
 \eqref{action} from the Einstein frame with metric 
 $\hat{g}_{\mu \nu}^E$ to the four-dimensional 
 string frame with metric $\hat{g}_{\mu \nu}^S$, where

\begin{equation}
\label{action3}
    \hat{g}_{\mu\nu}^E=e^{-2\varphi} \hat{g}_{\mu\nu}^S \Rightarrow
  \sqrt{\hat{g}^E}=\sqrt{\det(\hat{g}^S e^{-2\varphi})}=\sqrt{ e^{-8\varphi}\det(\hat{g}^S)}=e^{-4\varphi}\sqrt{\hat{g}^S} \;.
\end{equation}
To re-write the Einstein-Hilbert term, we use
the transformation rule of the Ricci scalar under a conformal rescaling, see for example \cite{Wald:1984rg}, appendix D.
After some manipulations and integration by parts, we obtain
the following string frame action:

\begin{align}
    \mathcal{S}_4&=\int d^4x\sqrt{\hat{g}_S}\Big[e^{-2\varphi}\Big(\hat{R}_S+4\hat{g}^{\mu\nu}_S\partial_{\mu}\varphi\partial_{\nu}\varphi\Big)-\frac{1}{2}\hat{g}^{\mu\rho}_S\hat{g}^{\nu\lambda}_S\hat{F}_{\mu\nu}\hat{F}_{\rho\lambda} \nonumber \\
    &-2e^{2\varphi}\left[\right.\partial^{\mu}\tilde{\varphi}+\frac{1}{2}(\zeta\partial^{\mu}\tilde{\zeta}-\tilde{\zeta}\partial^{\mu}\zeta)]\left[\right.\partial_{\mu}\tilde{\varphi}+\frac{1}{2}(\zeta\partial_{\mu}\tilde{\zeta}-\tilde{\zeta}\partial_{\mu}\zeta)] \nonumber \\
    &-\left(\partial_{\mu}\zeta\partial^{\mu}\zeta+\partial^{\rho}\tilde{\zeta}\partial_{\rho}\tilde{\zeta}\right)\Big] \label{action3a} \;.
\end{align}
Next, we restore the Kalb-Ramond B-field by Hodge-dualizing
the universal axion  $\tilde{\varphi}$. This procedure is standard, 
but we give some details in the appendix, in particular to be explicit about
our conventions.

\begin{align}
    \mathcal{S}_4&=\int d^4x\sqrt{\hat{g}_S}\Big[e^{-2\varphi}\Big(\hat{R}_S+4\hat{g}^{\mu\nu}_S\partial_{\mu}\varphi\partial_{\nu}\varphi\Big)-\frac{1}{2}\hat{g}^{\mu\rho}_S\hat{g}^{\nu\lambda}_S\hat{F}_{\mu\nu}\hat{F}_{\rho\lambda}\nonumber \\
    &-\frac{1}{12}e^{-2\varphi}\hat{H}_{\mu\nu\rho}\hat{H}^{\mu\nu\rho}-\frac{1}{6}(\zeta\partial^{\lambda}\tilde{\zeta}-\tilde{\zeta}\partial^{\lambda}\zeta)\epsilon^{\mu\nu\rho\lambda}\hat{H}_{\mu\nu\rho}\nonumber \\
    &-\left(\partial_{\mu}\zeta\partial^{\mu}\zeta+\partial^{\rho}\tilde{\zeta}\partial_{\rho}\tilde{\zeta}\right)\Big] \label{action5} \;.
\end{align}
While the Einstein frame action \eqref{action2} organsises the eight dynamical 
on-shell degrees of freedom into the bosonic components $\hat{g}_{\mu \nu}, \hat{A}_\mu$
of the $\mathcal{N}=2$ supergravity multiplet and the bosonic components $\varphi, \tilde{\varphi},
\zeta, \tilde{\zeta}$ of the universal hypermultiplet, in the string frame action 
\eqref{action5} these fields have been re-packaged into the NS-NS degrees of freedom 
$\hat{g}^S_{\mu \nu}$, $\hat{B}_{\mu \nu}$, $\varphi$ and the R-R degrees of
freedom $\hat{A}_\mu, \zeta,\tilde{\zeta}$.

We are now ready to perform the dimensional reduction. 
The coordinates are split into the coordinate $y$ along the direction we reduce over, and 
the remaining coordinates $x^\mu$, where $\mu$
from now on only takes three values. 
The four-dimensional line element
is decomposed according to 
\[
d\hat{s}_{4}^2 = e^{2\alpha \sigma} ds_3^2
+ e^{2\beta \sigma} (V+dy)^2\;,
\]
where $ds_3^2$ is the three-dimensional line
element, $V=V_\mu dx^\mu$ the Kaluza-Klein 
vector, $\sigma$ the Kaluza-Klein scalar,
and where $\alpha,\beta$ are numerical parameters.
For a string frame to string frame reduction
these parameter take the values $\alpha=0$, $\beta=1$ (in any dimension), so that
\[
d\hat{s}_{(4),(S)}^2 = ds_{(3),(S)}^2 + e^{2\sigma} (V+dy)^2 \;.
\]

Under dimensional reduction, part of the four-dimensional 
diffeomorphisms become three-dimensional $U(1)$ gauge transformations.
The associated gauge field is the Kaluza-Klein vector $V_\mu$. It is well known
that in order to obtain a dimensionally reduced Lagrangian which is manifestly 
invariant under three-dimensional gauge transformations, one needs to include
certain `shifts' proportional to the KK-vector $V=V_\mu dx^\mu$ when decomposing
four-dimensional into three-dimensional gauge fields. The four-dimensional
gauge field is decomposed as
$$\hat{A} =\xi dy+(A_{\mu} + \xi V_\mu) dx^{\mu} $$  into a three-dimensional gauge field $A_\mu$ and a scalar $\xi$.
The four-dimensional Kalb-Ramond field strength is decomposed 
as $\hat{H}_3 = H_3 + F' \wedge dy$ into
a three-dimensional three-form $H_3=dB_2$ and 
two-form $F'=dV'$. The corresponding two-form potential decomposes
as $\hat{B}_2 = B+ V' \wedge dy$. These expressions are re-arranged as
follows:
\begin{eqnarray*}
\hat{H}_3 &=& H_3 - F' \wedge V + F' \wedge ( dy + V    ) =: \tilde{H}_3 + F' \wedge ( dy + V    )\\
\hat{B}_2 &=& B_2 - V \wedge V' + V' \wedge (dy + V)\;.
\end{eqnarray*}
The modified three-dimensional Kalb-Ramond field strength
\[
\tilde{H}_3 := H_3 - F' \wedge V
\]
is invariant under $U(1)$ transformations associated with $V'$, since the 
transformation of $H_3=dB_2$ (induced by diffeomorphisms around the Kaluza-Klein 
circle) is compensated by the transformation of the `transgression term' $-F'\wedge V$.
While the scalars reduce trivially, we need to introduce the three-dimensional dilaton 
$$\bar{\varphi}:=\varphi-\frac{1}{2}\sigma$$ in order to obtain
the standard form of three-dimensional string frame
action
 \begin{align}
    \mathcal{S}_3=\int d^3x \sqrt{g_S}\Bigg(&e^{-2\bar{\varphi}}R_S+4e^{-2\bar{\varphi}}g_S^{\mu\nu}\partial_{\mu}\bar{\varphi}\partial_{\nu}\bar{\varphi}-e^{-2\bar{\varphi}}g_S^{\mu\nu}\partial_{\mu}\sigma\partial_{\nu}\sigma-\frac{1}{4}e^{-2(\bar{\varphi}-\sigma)}V_{\mu\nu}V^{\mu\nu} \nonumber \\
    &-\frac{1}{2}e^{\sigma}(F^{*}_{\mu\nu}+\xi V_{\mu\nu})(F^{*\mu\nu}+\xi V^{\mu\nu})-e^{-\sigma}\partial_{\mu}\xi\partial^{\mu}\xi \nonumber \\
    &-\frac{1}{12}e^{-2\bar{\varphi}}\tilde{H}_{\mu\nu\rho}\tilde{H}^{\mu\nu\rho}-\frac{1}{4}e^{-2(\bar{\varphi}+\sigma)}F'_{\mu\nu}F'^{\mu\nu} \nonumber\\
    &-\frac{1}{2}\epsilon^{\mu\nu y\lambda}F'_{\mu\nu}(\zeta\partial_{\lambda}\tilde{\zeta}-\tilde{\zeta}\partial_{\lambda}\zeta) \nonumber\\
    &-e^{\sigma}\left[\partial_{\mu}\zeta\partial^{\mu}\zeta+\partial^{\rho}\tilde{\zeta}\partial_{\rho}\tilde{\zeta}\right]\Bigg) \;.
\end{align}
The independent dynamical degrees of freedom are the three-dimensional dilaton $\bar{\varphi}$, the Kaluza-Klein scalar $\sigma$, the two R-R scalars $\zeta, \tilde{\zeta}$, the Kaluza-Klein gauge field $V_\mu$, the vector field $A_\mu$ 
and scalar field $\xi$
descending from the Maxwell field, and the vector $V'_\mu$ descending from the 
Kalb-Ramond field. The three-dimensional metric $g^S_{\mu \nu}$ and Kalb-Ramond 
field $B_{\mu \nu}$ do not carry local on-shell degrees of freedom. 

We expect T-duality to manifest itself as a symmetry 
of the three-dimensional action, where two groups of fields are exchanged for each
other. The present form of the action is not quite suitable for exhibiting this 
symmmetry, since the dynamical degrees of freedom are 5 scalars and 3 vector fields.
To obtain a symmetric expression, we dualize the reduced Maxwell field $F_{\mu\nu}$
into a scalar $\tilde{\xi}$.
We add the following Lagrange multiplier: 
\begin{equation}
    \mathcal{L}_m =-\epsilon^{\mu\nu\rho}F_{\mu\nu}\partial_{\rho}\tilde{\xi} \;.
\end{equation}
Then we eliminate $F_{\mu \nu}$ using its algebraic equation 
of motion
\begin{align}
        F^{\mu\nu}&=-e^{-\sigma}\epsilon^{\mu\nu\rho}\partial_{\rho}\tilde{\xi}-\xi V^{\mu\nu} \;.
\end{align}
After integration by parts we obtain the final form of the three-dimensional action,
\begin{align}
    \mathcal{S}_3=\int d^3x \sqrt{g_S}\Bigg(&e^{-2\bar{\varphi}}R_S+4e^{-2\bar{\varphi}}g_S^{\mu\nu}\partial_{\mu}\bar{\varphi}\partial_{\nu}\bar{\varphi}-e^{-2\bar{\varphi}}g_S^{\mu\nu}\partial_{\mu}\sigma\partial_{\nu}\sigma \nonumber \\&  -\frac{1}{4}e^{-2(\bar{\varphi}-\sigma)}V_{\mu\nu}V^{\mu\nu} -\frac{1}{4}e^{-2(\bar{\varphi}+\sigma)}F'_{\mu\nu}F'^{\mu\nu}\nonumber\\
    &-e^{-\sigma}\left[\partial_{\mu}\xi\partial^{\mu}\xi+\partial^{\rho}\tilde{\xi}\partial_{\rho}\tilde{\xi}\right] - e^{\sigma}\left[\partial_{\mu}\zeta\partial^{\mu}\zeta+\partial^{\rho}\tilde{\zeta}\partial_{\rho}\tilde{\zeta}\right] \nonumber  \\
&    +\epsilon^{\mu\nu\rho}\xi V_{\mu\nu}\partial_{\rho}\tilde{\xi}
 -\epsilon^{\mu\nu y \lambda}F'_{\mu\nu}\zeta\partial_{\lambda}\tilde{\zeta} 
\nonumber\\
    &-\frac{1}{12}e^{-2\bar{\varphi}}\tilde{H}_{\mu\nu\rho}\tilde{H}^{\mu\nu\rho}
\Bigg) \;, 
\label{action6}
\end{align}
whose dynamical fields are the six scalars 
$\bar{\varphi}, \sigma, \xi, \tilde{\xi}, \zeta, \tilde{\zeta}$
and two gauge fields $V_{\mu \nu},F'_{\mu \nu}$. 
This action
is invariant under the following non-trivial involutive field transformation:
\begin{equation}
\label{T-duality_3d}
    \bar{\varphi}\rightarrow\bar{\varphi}, \sigma\rightarrow-\sigma, 
    \tilde{\xi} \leftrightarrow\tilde{\zeta}, \xi \leftrightarrow\zeta, V'_{\mu}\leftrightarrow -V_{\mu}, B_2 \rightarrow B_2- V' \wedge V \;.
\end{equation}

Since $\sigma \rightarrow -\sigma$ amounts to inverting the radius of the internal circle in string units, this $\mathbb{Z}_2$-symmetry 
is indeed T-duality. Using the relation between four- and three-dimensional
fields, we can lift the three-dimensional transformation to 
four dimensions and obtain a four-dimensional version of the
Buscher rules. We find that for the four-dimensional NS-NS 
fields, that is, for the four-dimensional dilaton, Kalb-Ramond form
and string frame metric, the Buscher rules take exactly the same form as in 
ten dimensions. For example, the standard transformation rule
for the four-dimensional dilaton follows from the invariance
of the three-dimensional dilaton and the transformation of the
KK-scalar:
\begin{equation}
    \varphi'=\bar{\varphi'}+\frac{1}{2}\sigma'\rightarrow \bar{\varphi}-\frac{1}{2}\sigma=\varphi-\frac{1}{2}\sigma-\frac{1}{2}\sigma=\varphi-\sigma=\varphi -\frac{1}{2}\ln{|\hat{g}_{yy}|} \;.
\end{equation}
We also obtain Buscher rules for the four-dimensional R-R fields $\hat{A}_\mu$, 
$\zeta$ and $\tilde{\zeta}$.
We see easily that
\begin{equation}
    \zeta \rightarrow \xi=\hat{A}_y \;.
\end{equation}
For the derivative of the scalar field $\tilde{\zeta}$ 
we obtain
\begin{equation}
    \partial_\mu \zeta \rightarrow \partial_\mu \tilde{\zeta} = 
    \frac{1}{2}\hat{\epsilon}_{y\mu\nu\lambda}\sqrt{\hat{g}_{yy}}\left[\hat{F}^{\mu\nu}+\hat{A}_{y}\left(\partial^{\mu}\left(\frac{\hat{g}^{\nu y}}{\hat{g}_{yy}}\right)-\partial^{\nu}\left(\frac{\hat{g}^{\mu y}}{\hat{g}_{yy}}\right)\right)\right] \;,
\end{equation}
see the appendix for details.

Note that there is no need to obtain a transformation
law for $\tilde{\zeta}$ itself. As is well known \cite{Ferrara:1989ik,Cortes:2015wca}, the hypermultiplet manifold
obtained by dimensional reduction of a theory of $n$  vector multiplets coupled to $\mathcal{N}=2$ supergravity has an isometry group which contains the Iwasawa subgroup of $\mathrm{SU}(1,n+2)$, acting
by affine transformations on the scalar fields. 
While we work in the
string frame, and have not dualized all vector fields into scalars,
the effect of dualization is to convert axionic shift transformations into gauge transformations, so that symmetries are preserved.\\

Similarly, we obtain the transformation for the field strength
of the Maxwell field:
\begin{equation}
\hat{F}_{\mu \nu} \rightarrow \hat{F}'_{\mu \mu } =  -\sqrt{\hat{g}_{yy}}\hat{\epsilon}^{y\mu\nu\rho}\partial_{\rho}\tilde{\zeta}+\zeta\left(\partial^{\mu}(\hat{B}_{\nu y})-\partial^{\nu}(\hat{B}_{\mu y})\right) \;,  
\end{equation}
see again the appendix for details.

To conclude this section, let us summarize the
four-dimensional Buscher rules that we have derived.
\begin{framed}
NS-NS sector:
\begin{align}
    &g'_{yy}=\frac{1}{g_{yy}}\;,\\
    &g'_{y\mu}=\frac{B_{y\mu}}{g_{yy}}\;, \\
    &g'_{\mu\nu}= g_{\mu\nu}+\frac{B_{y\mu}B_{y\nu}-g_{y\mu}g_{y\nu}}{g_{yy}}\\
    &B_{y\mu}=\frac{g_{y\mu}}{g_{yy}}\;, \\
    &B'_{\mu\nu}=B_{\mu\nu}+\frac{g_{y\mu}B_{y\nu}-B_{y\mu}g_{y\nu}}{g_{yy}}\;, \\
    &\varphi'=\varphi-\frac{1}{2}\ln |g_{yy}| \;.
\end{align}
\end{framed}
\begin{framed}
R-R sector:    
\begin{align}
    &\zeta'=\hat{A}_y \;, \\
    &\partial_{\lambda}\tilde{\zeta}'=\frac{1}{2}\hat{\epsilon}_{y\mu\nu\lambda}\sqrt{|\hat{g}_{yy}|}\left[\hat{F}^{\mu\nu}+\hat{A}_{y}\left(\partial^{\mu}\left(\frac{\hat{g}^{\nu y}}{\hat{g}_{yy}}\right)-\partial^{\nu}\left(\frac{\hat{g}^{\mu y}}{\hat{g}_{yy}}\right)\right)\right] \;, \\
    &\hat{A'}_y=\zeta \;, \\
    &\hat{F'}^{\mu\nu}=-\sqrt{|\hat{g}_{yy}|}\hat{\epsilon}^{y\mu\nu\rho}\partial_{\rho}\tilde{\zeta}+\zeta\left(\partial^{\mu}(\hat{B}_{\nu y})-\partial^{\nu}(\hat{B}_{\mu y})\right) \;.
\end{align}
\end{framed}

\subsection{Generalization to the twisted model and to timelike reduction}

So far we have formulated the standard space-like 
T-duality for the EM-UHM model in the
string frame and obtained the explicit relation between
the fields. 
There are two modifications of this procedure which lead to
relative sign flips. Firstly, we can change the starting
point and apply a space-like T-duality to the twisted version
of the theory, where the kinetic terms of the R-R fields
have been flipped. Starting from the Einstein frame action 
obtained in \cite{Medevielle:2021wyx}, 
\begin{align}
    \mathcal{S}_4&=\int d^4x\sqrt{\hat{g}_E}\Big(\hat{R}_E\color{red}+\color{black}\frac{1}{2}\hat{g}_E^{\mu\rho}\hat{g}_E^{\nu\lambda}\hat{F}_{\mu\nu}\hat{F}_{\rho\lambda}-2\hat{g}_E^{\mu\nu}\partial_{\mu}\varphi\partial_{\nu}\varphi  \nonumber \\
    &-2e^{4\varphi}\left[\right.\partial^{\mu}\tilde{\varphi}+\frac{1}{2}(\zeta\partial^{\mu}\tilde{\zeta}-\tilde{\zeta}\partial^{\mu}\zeta)]\left[\right. \partial_{\mu}\tilde{\varphi}+\frac{1}{2}(\zeta\partial_{\mu}\tilde{\zeta}-\tilde{\zeta}\partial_{\mu}\zeta)] \nonumber\\
    &\color{red}+\color{black}e^{2\varphi}\left[\partial_{\mu}\zeta\partial^{\mu}\zeta+\partial^{\rho}\tilde{\zeta}\partial_{\rho}\tilde{\zeta}\right]\Big) \;,
\end{align}
conversion to the string frame 
\begin{align}
    \mathcal{S}_4&=\int d^4x\sqrt{\hat{g}_S}\Big[e^{-2\varphi}\Big(\hat{R}_S+4\hat{g}^{\mu\nu}_S\partial_{\mu}\varphi\partial_{\nu}\varphi\Big)\color{red}+\color{black}\frac{1}{2}\hat{g}^{\mu\rho}_S\hat{g}^{\nu\lambda}_S\hat{F}_{\mu\nu}\hat{F}_{\rho\lambda}\nonumber \\
    &-\frac{1}{12}e^{-2\varphi}\hat{H}_{\mu\nu\rho}\hat{H}^{\mu\nu\rho}-\frac{1}{6}(\zeta\partial^{\lambda}\tilde{\zeta}-\tilde{\zeta}\partial^{\lambda}\zeta)\epsilon^{\mu\nu\rho\lambda}\hat{H}_{\mu\nu\rho}\nonumber \\
    &\color{red}+\color{black}\left(\partial_{\mu}\zeta\partial^{\mu}\zeta+\partial^{\rho}\tilde{\zeta}\partial_{\rho}\tilde{\zeta}\right)\Big] \label{action5bis} \;,
\end{align}
followed by reduction over space gives
\begin{align}
    \mathcal{S}_3=\int d^3x \sqrt{g_S}\Bigg(&e^{-2\bar{\varphi}}R_S+4e^{-2\bar{\varphi}}g_S^{\mu\nu}\partial_{\mu}\bar{\varphi}\partial_{\nu}\bar{\varphi}-e^{-2\bar{\varphi}}g_S^{\mu\nu}\partial_{\mu}\sigma\partial_{\nu}\sigma-\frac{1}{4}e^{-2(\bar{\varphi}-\sigma)}V_{\mu\nu}V^{\mu\nu}  \nonumber \\
&    \color{red}+\color{black}e^{-\sigma}\left[\partial_{\mu}\xi\partial^{\mu}\xi+\partial^{\rho}\tilde{\xi}\partial_{\rho}\tilde{\xi}\right]
    +\epsilon^{\mu\nu\rho}\xi V_{\mu\nu}\partial_{\rho}\tilde{\xi}
    -\frac{1}{12}e^{-2\varphi}\tilde{H}_{\mu\nu\rho}\tilde{H}^{\mu\nu\rho} \nonumber \\ &-\frac{1}{4}e^{-2(\bar{\varphi}+\sigma)}F'_{\mu\nu}F'^{\mu\nu}
    -\epsilon^{\mu\nu y \lambda}F'_{\mu\nu}\zeta\partial_{\lambda}\tilde{\zeta} \color{red}+\color{black}e^{\sigma}\left[\partial_{\mu}\zeta\partial^{\mu}\zeta+\partial^{\rho}\tilde{\zeta}\partial_{\rho}\tilde{\zeta}\right]\Bigg)
\end{align}
This is again invariant under \eqref{T-duality_3d}, so that
lifting back over space after the transformation gives back
the original Langrangian. This is as expected from the Einstein frame 
point of view, since  we know from \cite{Medevielle:2021wyx} that the three-dimensional Einstein frame theory has the scalar manifold 
\[
\mathcal{N}'' = \frac{U(1,2)}{U(1,1) \times U(1)} \times \frac{U(1,2)}{U(1,1)\times U(1)}\;.
\]
This has two isomorphic factors which
get exchanged by T-duality. Thus the two versions of
the EM-UHM theory get mapped to themselves under spacelike
T-duality. 

The second modification is to perform a timelike T-duality, 
that is to reduce over time, perform an involutive field 
transformation, and lift back over time. Starting from the
standard EM-UHM theory, reduction over time gives
\begin{align}
    \mathcal{S}_3=\int d^3x \sqrt{g_S}\Bigg(&e^{-2\bar{\varphi}}R_S+4e^{-2\bar{\varphi}}g_S^{\mu\nu}\partial_{\mu}\bar{\varphi}\partial_{\nu}\bar{\varphi}-e^{-2\bar{\varphi}}g_S^{\mu\nu}\partial_{\mu}\sigma\partial_{\nu}\sigma
    \nonumber \\
    &\color{red}+\color{black}\frac{1}{4}e^{-2(\bar{\varphi}-\sigma)}V_{\mu\nu}V^{\mu\nu} 
    \color{red}+\color{black}\frac{1}{4}e^{-2(\bar{\varphi}+\sigma)}F'_{\mu\nu}F'^{\mu\nu} 
    \nonumber \\
    &\color{red}+\color{black}e^{-\sigma}\left[\partial_{\mu}\xi\partial^{\mu}\xi+\partial^{\rho}\tilde{\xi}\partial_{\rho}\tilde{\xi}\right] 
    -e^{\sigma}\left[\partial_{\mu}\zeta\partial^{\mu}\zeta+\partial^{\rho}\tilde{\zeta}\partial_{\rho}\tilde{\zeta}\right]
\nonumber\\
    &+\epsilon^{\mu\nu\rho}\xi V_{\mu\nu}\partial_{\rho}\tilde{\xi}
        -\epsilon^{\mu\nu y \lambda}F'_{\mu\nu}\zeta\partial_{\lambda}\tilde{\zeta}\nonumber \\
        &-\frac{1}{12}e^{-2\varphi}\tilde{H}_{\mu\nu\rho}\tilde{H}^{\mu\nu\rho}
    \Bigg) \;.
\end{align}
The reduction of the twisted EM-UHM theory over time gives
\begin{align}
    \mathcal{S}_3=\int d^3x \sqrt{g_S}\Bigg(&e^{-2\bar{\varphi}}R_S+4e^{-2\bar{\varphi}}g_S^{\mu\nu}\partial_{\mu}\bar{\varphi}\partial_{\nu}\bar{\varphi}-e^{-2\bar{\varphi}}g_S^{\mu\nu}\partial_{\mu}\sigma\partial_{\nu}\sigma 
    \nonumber \\
&    \color{red}+\color{black}\frac{1}{4}e^{-2(\bar{\varphi}-\sigma)}V_{\mu\nu}V^{\mu\nu}
\color{red}+\color{black}\frac{1}{4}e^{-2(\bar{\varphi}+\sigma)}F'_{\mu\nu}F'^{\mu\nu} \nonumber
\\
    &-e^{-\sigma}\left[\partial_{\mu}\xi\partial^{\mu}\xi+\partial^{\rho}\tilde{\xi}\partial_{\rho}\tilde{\xi}\right]
\color{red}+\color{black}e^{\sigma}\left[\partial_{\mu}\zeta\partial^{\mu}\zeta+\partial^{\rho}\tilde{\zeta}\partial_{\rho}\tilde{\zeta}\right] \nonumber
     \\
    &+\epsilon^{\mu\nu\rho}\xi V_{\mu\nu}\partial_{\rho}\tilde{\xi}
    -\epsilon^{\mu\nu y \lambda}F'_{\mu\nu}\zeta\partial_{\lambda}\tilde{\zeta}
  \nonumber  \\
    &-\frac{1}{12}e^{-2\varphi}\tilde{H}_{\mu\nu\rho}\tilde{H}^{\mu\nu\rho}
        \Bigg) \;.
        \end{align}
Since these two actions get mapped to each other under \eqref{T-duality_3d},
timelike T-duality maps the standard EM-UHM theory to the
twisted one, and vice versa. This is again as expected,
since the time-like reduction of both theories gives, in the Einstein frame, 
a three-dimensional Euclidean theory with a scalar manifold \eqref{N-prime}
which has two non-isometric factors.

By tracing the effects of sign flips compared to the previous
section, we obtain Buscher rules for all cases. It turns out that
the rules for the NS-NS fields are universal. The rules for the
R-R fields take the form
\begin{framed}
\begin{align}
    &\zeta'=\hat{A}_y  \label{RR1}\\
    &\hat{A'}_y=\zeta \label{RR2}\\
    &\partial_{\lambda}\tilde{\zeta}'=\alpha_1\frac{1}{2}\hat{\epsilon}_{y\mu\nu\lambda}\sqrt{\hat{g}_{yy}}\left[\hat{F}^{\mu\nu}+\hat{A}_{y}\left(\partial^{\mu}\left(\frac{\hat{g}^{\nu y}}{\hat{g}_{yy}}\right)-\partial^{\nu}\left(\frac{\hat{g}^{\mu y}}{\hat{g}_{yy}}\right)\right)\right] \label{RR3}\\
    &\hat{F'}^{\mu\nu}=-\alpha_2\sqrt{\hat{g}_{yy}}\hat{\epsilon}^{y\mu\nu\rho}\partial_{\rho}\tilde{\zeta}+\zeta\left(\partial^{\mu}(\hat{B}_{\nu y})-\partial^{\nu}(\hat{B}_{\mu y})\right)  \label{RR4}\\
    &     \alpha_1 = \left\{ \begin{array}{ll} 
    +1 & \mbox{if original theory untwisted,} \\
    -1 & \mbox{if original theory twisted,} \\
    \end{array} \right. \,\;\;\;
    \alpha_2 = \left\{ \begin{array}{ll} 
    +1 & \mbox{if dual theory untwisted,} \\
    -1 & \mbox{if dual  theory twisted.} \\
    \end{array} \right.  \nonumber
\end{align}    
\end{framed}

\section{Dualizing a cosmological solution (from type-IIA to type-IIB*/IIB)}

\subsection{The dual solution}

With the Buscher rules in place, we can now T-dualize the 
cosmological solution \eqref{solution} of the EM-UHM model
\eqref{action}, \eqref{action5}, which we recall for reference.
Since all quantities are four-dimensional, we omit `hats' in this section. The 
indices $\mu, \nu$ take values $0,1,2,3$, equivalently $t,r,x,y$ (where the direction indexed by $0$ or $t$ is timelike for $0<r<r_*$). 
The metric is
\begin{equation}
ds^2 = - f(r) dt^2 + \frac{1}{f(r)}dr^2 + r^2 (dx^2 + dy^2) \;,\;\;    
f(r) = - \frac{2M}{r} + \frac{e^2}{r^2} \;,
\end{equation}
while $B$-field and dilaton are trivial and have been set to zero.
The Einstein frame coincides with the string frame. 
Since the only non-trivial fields are the metric and the Maxwell field, this is a solution to Einstein-Maxwell theory. The 
mass-like parameter $M$ and electric charge $e$ have 
to satisfy $M>0$ and $M^2>e^2>0$ to avoid naked singularities.
For these values there is a Killing horizon at $r=r_* =\frac{e^2}{2M}$, which separates a static region $0<r<r_*$ from 
a non-static region $r>r_*$. The solution has a curvature singularity for $r\rightarrow 0+$ and is asymptotic to a 
Kasner cosmological solution for $r\rightarrow \infty$,
see \cite{Gutowski:2019iyo,Gutowski:2020fzb} for details.

We T-dualize with 
respect to the Killing vector field $\xi=\partial_t$ which is timelike
for $0<r<r_*=\frac{e^2}{2M}$ and spacelike for $r_* < r < \infty$. Applying
the NS-NS Buscher rules, the T-dual string frame metric is
\begin{equation}
\label{IIBstar_stringframe}
    ds'^2_{\mathrm{S}}=\frac{- dt^2 + dr^2 }{f(r)}+r^2(dx^2+dy^2) \;,
\end{equation}
for all positive $r\not=r_*$. Since the initial metric is diagonal, 
the $B$-field remains trivial. However,  in the dual solution we have 
a non-constant dilaton 
\begin{equation}
    \varphi'=-\frac{1}{2}\ln|f(r)| = -\frac{1}{2} \ln \left| 
    -\frac{2M}{r} + \frac{e^2}{r^2} \right|,
\end{equation}
and therefore the dual string frame and Einstein frame metric differ
by a conformal factor. 

Besides the metric, the only non-trivial field in the original solution 
is the Maxwell field
\begin{equation}
\label{solution2}
F = - \frac{e}{r^2} dt \wedge dr \Leftarrow A=-\frac{e}{r}dt \;,
\end{equation}
where we have made a choice for the gauge potential, which
is only determined up to an additive constant.

For the Maxwell field
we have to apply the R-R Buscher rules, which distinguish between
the standard and the twisted EM-UHM model. The initial model is untwisted, $\alpha_1=1$, while
for the dual models this depends on the value of $r$. For $r<r_*$, 
T-duality is timelike, $\alpha_2=-1$, while for $r>r_*$ it is  spacelike,
$\alpha_2=1$.
From \eqref{RR1},\eqref{RR2},\eqref{RR4} we immediately find
\begin{equation}
      \zeta'=-\frac{e}{r}\;, \;\;\; A'_{y}=0\quad \text{and}\quad F'_{\mu\nu}=0
      \;.
      \end{equation}
Moreover, since we dualize over the direction $y=t$, and since the metric
is diagonal, \eqref{RR3} implies
\begin{equation}
     \partial_{\lambda}\tilde{\zeta}'=0 \;.  
\end{equation}
The freedom of choosing a constant value for $\tilde{\zeta}'$ reflects the
freedom to perform gauge transformations on the Maxwell potential $A_\mu$.
Since the
parameter $\alpha_2$ does not enter into the solution, the solution takes
the same form for $r<r_*$ and $r>r_*$. While it is not required (since T-duality is a solution generating
transformation), it is straightforward to check that the dual 
solution satisfies the equations of motion of the twisted and of the standard
EM-UHM model for $r<r_*$ and $r>r_*$, respectively.

The non-trivial fields in the T-dual solution are the string frame metric $g'^{(S)}_{\mu \nu}$, the dilaton $\varphi'$ and the axion $\zeta'$. Truncating the actions
\eqref{action5}, \eqref{action5bis} to these fields we obtain a dilaton-axion
system coupled to gravity, with string frame action
\begin{align}
    \mathcal{S}_4&=\int d^4x\sqrt{{g}_S}\Big[e^{-2\varphi'}\Big({R}_S+4{g}^{\mu\nu}_S\partial_{\mu}\varphi'\partial_{\nu}\varphi'\Big)  \color{red}\mp \color{black} g^{\mu \nu}_S \partial_{\mu}\zeta'\partial_{\nu}\zeta' \Big]\;.
    \label{action7}
\end{align}
The Einstein frame version of this action
\[
\mathcal{S}_4 = \int d^4x \sqrt{g_E} \left( R_E - 2 g^E_{\mu \nu} \partial_\mu
\varphi' \partial_\nu \varphi' \color{red} \mp \color{black} e^{2\varphi'} g^E_{\mu \nu}
\partial_\mu \zeta' \partial_\nu \zeta'\right) \;,
\]
is a nonlinear sigma model with target space $\mathrm{SL}(2,\mathbb{R})/\mathrm{SO}(2)$
and $\mathrm{SL}(2,\mathbb{R})/\mathrm{SO}(1,1)$, respectively, coupled to 
gravity. The lower (upper) sign applies for timelike (spacelike) T-duality, thus
establishing the relations we have provided in Figures \ref{fig:1} and \ref{fig:2}.

The sign flip in the dilaton-axion action implies that the fields
$(\varphi', \zeta')$ have to `move' 
from one target space to the other.  We will see below
that the scalars run off to a boundary of their respective
scalar manifolds when approaching $r=r_*$. With our embedding
of the cosmological solution into type-IIA string theory, 
$r=r_*$ is an interface between its timelike and spacelike
T-duals, type-IIB$^*$ for $r<r_*$ and type-IIB for $r>r_*$.

\subsection{Behaviour of the T-dual solution at special points}

\begin{figure}[h]
\centering
\includegraphics[width=0.5\textwidth]{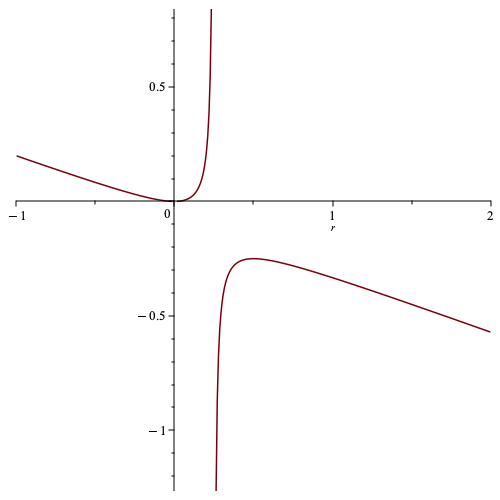}
\caption{Graph of the function $h(r)=1/f(r)$ with parameter values $M=2$, $e=1$. The shown
behaviour applies to any choice of parameters where $M>0$ and $M^2>e^2$.
The points $r=0$ and $r=r_* = 0.25$ on the horizontal axis 
correspond to the T-duality images of the singularity and of the horizon of the cosmological solution. We have displayed the graph 
for negative values $r<0$ to discuss a possible extension of the solution to these values. 
\label{Fig:f_extended_range}}
\end{figure}

We will now take a closer look at the behaviour of the dual solution
at the image of the singularity $r=0$ and at the image of 
the horizon $r=r_*$. In Figure \ref{Fig:f_extended_range} we display the
graph of the function $h(r):= \frac{1}{f(r)}$, 
which is extremely useful for understanding the behaviour of the solution. 
While specific values for $e,M$ have been used in this graph, it is straightforward to show analytically that 
the qualitative behaviour shown there is universal for $M>|e|>0$. The 
function $h=1/f$ has a pole at  $r=r_*$ and is analytic otherwise. It is negative 
for $r>r_*$ with a local maximum at $r=2r_*$ and 
non-negative for $r<r_*$ with a unique zero (which therefore
is the unique local minimum) at $r=0$. It approaches $\pm \infty$
for $r\rightarrow \mp \infty$ and $r\rightarrow r_* \mp 0$, as 
indicated in the graph.

At $r=0$ all string frame metric coefficients go
to zero, so that the metric completely degenerates (it has rank zero as a matrix):
\[
g^{(S)}_{\mu \nu} \xrightarrow[r\rightarrow 0]{} 0 \;.
\]
At the image $r=r_*$ of the horizon $f(r)$ has its zero, so that $h(r)$ and, hence, $g_{tt}^{(S)}$ and $g_{rr}^{(S)}$ diverge:
\[
g^{(S)}_{tt} \xrightarrow[r \rightarrow r_* \mp ]{} \pm \infty \;,\;\;
g^{(S)}_{rr} \xrightarrow[r \rightarrow r_* \mp ]{} \mp \infty \;.
\]
The Ricci tensor of the string frame metric
is
\begin{eqnarray}
    R^{(S)}_{\mu\nu}=\text{diag} && \bigg(\frac{e^4-2e^2Mr+2M^2r^2}{r^2(e^2-2Mr)^2},\frac{3e^4-10e^2Mr+6M^2r^2}{r^2(e^2-2Mr)^2},  \nonumber\\
    && -\frac{e^2-2Mr}{r^2},-\frac{e^2-2Mr}{r^2} \bigg) \;,
\end{eqnarray}
and the corresponding Ricci scalar is:
\begin{equation}
    R^{(S)}=\frac{4M^2}{r^2(-e^2+2Mr)} \;.
\end{equation}
The string frame metric has singular Ricci curvature at 
both $r=0$ and $r=r_*$. 

\begin{figure}[h]
\centering
\includegraphics[width=0.5\textwidth]{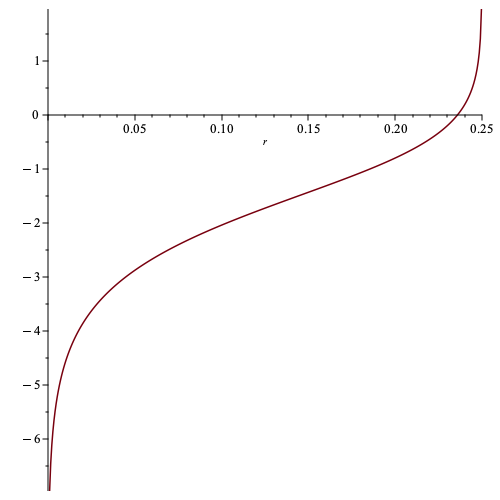}
\caption{Graph of dilaton $\varphi'(r)$ for the range $0<r<r_*$. 
The T-dual of the singularity and of the horzion are located
at $r=0$ and $r=r_*=0.25$, respectively. 
The behaviour shown by this graph
is universal for $M>0$, $M^2>e^2$.
\label{Fig:Dilaton_restricted_range}}
\end{figure}

\begin{figure}[h]
\centering
\includegraphics[width=0.5\textwidth]{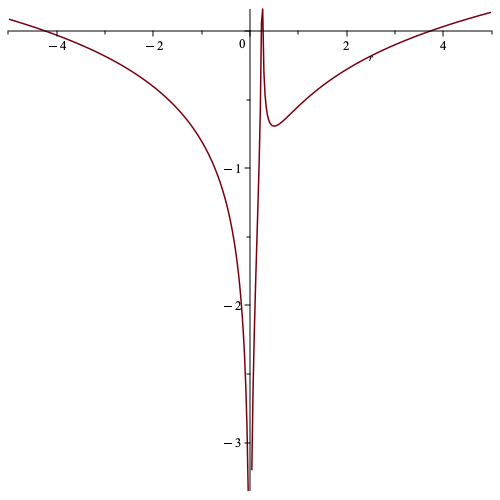}
\caption{Graph of the dilaton $\varphi'(r)$ for an extended which displays its four zeros. The behaviour shown by this
graph is universal for $M>0$, $M^2>e^2$.\label{Fig:Dilaton_extended_range}}
\end{figure}

Next, let us look at the two scalars. The axion $\zeta' = -e/r$
becomes singular at $r\rightarrow 0$ but is regular at $r=r_*$.
The dilaton 
\begin{equation}
    \varphi'=-\frac{1}{2}\ln \left|f(r)\right|=-\frac{1}{2}\ln\left|\frac{-2M}{r}+\frac{e^2}{r^2}\right|
\end{equation}
has a more complicated profile. It becomes infinite at the T-dual images 
of the singularity and of the horizon:
\begin{equation}
    \varphi'  \xrightarrow[r \rightarrow 0\pm]{}  - \infty \;,\;\;
    \varphi'  \xrightarrow[r \rightarrow r_*\pm ]{}  \infty \;, \;\;
\end{equation}
The dilaton $\varphi'$ has four zeros
\begin{equation}
    \varphi'(r) = 0 \Rightarrow f(r) = -\frac{2M}{r} + \frac{e^2}{r^2} = 
    \pm 1 \Rightarrow r = - M \pm \sqrt{M^2+ e^2} \;, 
    M \pm\sqrt{M^2- e^2} \;.
\end{equation}
One of the zeros, $r_0 = - \sqrt{M^2+e^2} - M < 0$ is `on the other side'
of the singularity $r=0$, while the other three zeros 
occur for positive $r>0$. Since
$\varphi'$ is continuous between $r=0$ and $r=r_*$ and approaches 
$\pm \infty$ at these points, it must have  an odd number of zeros between these values.
Since $r=r_*$ is the unique zero of $f(r)$, two of the zero of $\varphi'$
must be on either side of $r_*$. Thus $\varphi'$ has one zero between 
$0$ and $r_*$ and two zeros between $r_*$ and $\infty$.
For illustration, we display the graph of $\varphi'$ for the image of the
interior solution, $0<r<r_*$ in Figure \ref{Fig:Dilaton_restricted_range} and as well
for an extended range which displays all zeros and singularities in 
Figure \ref{Fig:Dilaton_extended_range}.

Due to the non-constant dilaton, the Einstein frame metric behaves differently from the string frame metric. Transforming \eqref{IIBstar_stringframe} to the Einstein frame, we obtain
\begin{eqnarray}
    ds'^2_E&=& e^{-2\varphi'} ds'^2_S = |f(r)| ds'^2_S  \\ &=& \left\{ 
    \begin{array}{ll}
    -dt^2+dr^2+(e^2-2Mr)(dx^2+dy^2)\;, & 0<r<r_* \;,\\
    + dt^2 - dr^2 -(e^2 -2Mr) (dx^2+dy^2)\;, & r_* < r < \infty \;.
    \end{array} \right. \nonumber
\end{eqnarray}
At the image of the singularity, $r=0$, all metric coefficients are finite, 
and can be continued analytically to $-\infty < r < r_*$. At $r=r_*$ the
metric coefficients $g^{(E)}_{tt}, g^{(E)}_{rr}$ have a finite discontinuity and change sign,
while $g^{(E)}_{xx}, g^{(E)}_{yy}$ have a first order zero and change sign without discontinuity. The signature of the metric is the
same for $r<r_*$ and $r>r_*$.

The Einstein frame Ricci tensor and
Ricci scalar are:
\begin{equation}
    R^{(E)}_{\mu\nu}=\pm \frac{2M^2}{(e^2-2Mr)^2}\delta_{rr}
\end{equation}
and
\begin{equation}
    R^{(E)}=\pm \frac{2M^2}{(e^2-2Mr)^2} \;,
\end{equation}
where the upper (lower) sign refers to $r<r_*$ ($r>r_*$). The Einstein frame 
Ricci curvature is finite at $r=0$, as it of course must be, since
all metric coefficients are analytical and the metric is extendible to negative
$r$. This means that the conformal structure of the T-dual solution is extendible to $r\leq 0$, since this only requires that there is a representative of the conformal class which can be extended. However, the conformal factor relating Einstein and string frame is the dilaton, which is a dynamical field of our theory. Therefore the T-dual solution as a whole is not extendible to $r\leq 0$, since the dilaton and the axion diverge for $r\rightarrow 0+$. At the image of the horizon, the Einstein frame Ricci curvature diverges, like
it does in the string frame. 

\subsection{Probing the singular points with geodesics}

Since the T-dual solution is singular at $r=0,r_*$ we now investigate whether these singularities are at finite or infinite `distance.' There are different meaningful concepts of distance that we can apply: we have two different conformal frames, and we can
probe the singularity with either spacelike or with null
geodesics. For the scalar fields we also have the concept of `distance in moduli
space,' to which we will come back in a later section.

We start by computing the length of a transverse spacelike 
geodesic connecting a point at position $r=R$ to a point at
$r=0$. Here `transverse' means constant $x,y$ coordinates, so that
we can set $dx=dy=0$ in the line element.
In the Einstein frame,
transverse curves locally see a two-dimensional Minkowski metric, with a
discontinuity at $r=r_*$:
\[
ds'^2_{(E)\mathrm{transv.}} = \pm \left( -dt^2 + dr^2 \right) \;.
\]
Therefore any point at $0< R<r_*$ can be connected to $r=0$ and to 
$r=r_*$ by a spacelike geodesic of finite length. This becomes more interesting in the string frame, 
where the transverse metric is
\[
ds'^2_{(S)\mathrm{transv.}} = h(r) \left( -dt^2 + dr^2 \right) \;,\; h(r) = 
\frac{1}{f(r)} = \frac{1}{-\frac{2M}{r} + \frac{e^2}{r^2}}  \;.
\]
We choose a point at transverse position $R$ in the interior region, 
$0<R<r_*$. Its spacelike distance to $r=0$ is given by the length of the curve
which connects $r=R$ to $r=0$ 
at constant $t,x,y$. 
Along such transverse spacelike geodesics, the line element further reduces to
\begin{equation}
    ds'^2_{(S)SL}=h(r) dr^2 \;.
\end{equation}
Choosing the geodesic length as curve parameter
$\lambda$ we obtain
\begin{equation}
h(r) \left(\frac{dr}{d\lambda}\right)^2=+1 \;.
\end{equation}
This can be solved for the infinitesimal geodesic length:
\begin{equation}
    d\lambda=\pm \frac{dr}{\sqrt{h}}=\pm \frac{dr}{\sqrt{-2M/r+e^2/r^2}} \;,
\end{equation}
where the upper sign corresponds to outgoing and the lower sign to ingoing
curves. We compute the length of an ingoing curve, thus obtaining
the spacelike (SL) string frame (SF) distance between $r=R$ and $r=0$:
\begin{eqnarray}
D_{\mathrm{SL,SR}}(0,R)
    &=&-\int^{0}_R\frac{dr}{\sqrt{-2M/r+e^2/r^2}}=\left[\frac{\sqrt{e^2-2Mr}(Mr+e^2)}{3M^2}\right]_R^0 \nonumber \\
    &=& \frac{e^2\sqrt{e^2}}{3M^2}-\left(\frac{\sqrt{e^2-2MR}(e^2+MR)}{3M^2}\right) \nonumber \\
     &=& \frac{e^2\sqrt{e^2}}{3 M^2} \left( 1 - (1+\frac{RM}{e^2})
     \sqrt{1 - \frac{2MR}{e^2}}\right) \;.
\end{eqnarray}
This expression is finite for $0<R<r_*$.\footnote{By construction, it is also positive.
This is straightforward to check by expanding the square root.} Moreover, the distance remains finite
for $R=r_*$, where it takes its maximal value (within the interval $[0,r_*]$):
\[
D_{SL,SR}(0,r_*) = \frac{e^2\sqrt{e^2}}{3M^2} \;.
\]
It follows without need for a separate computation that the spacelike distance 
from any point at $r=R$, with $0<R<r_*$ to $r=r_*$ is finite. Both singularities are
at finite spacelike distance from any interior point.

For null geodesics there is no concept of distance, but 
the distinguished class of affine curve parameters. 
We will use an affine curve parameter as substitute for 
the length, but since such parameters are only unique
up to affine transformations
the only relevant information is whether this `affine length'
is zero, finite, or infinite. 
Consider the transverse null curves 
\begin{equation}
    x^{\mu}(r)=(-r+R,r,0,0) \quad\text{and}\quad \dot{x}^{\mu}=(-1,1,0,0) \;,
\end{equation}
with curve parameter $r$, 
which start at time $t=0$ at transverse position $r=R$ and propagate inwards.
The non vanishing Christoffel symbols of the line element are
\begin{equation}
    \Gamma^t_{rt}=\frac{1}{2}h^{-1}\dot{h}, \quad \Gamma^x_{rx}=r^{-1}, \quad \Gamma^r_{tt}=\frac{1}{2}h^{-1}\dot{h},\quad \Gamma^r_{rr}=\frac{1}{2}h^{-1}\dot{h},\quad \Gamma^r_{xx}=\frac{1}{2}h^{-1}(-2r).
\end{equation}
We compute the absolute covariant derivatives 
\[(
\nabla_{\dot{x}} \dot{x})^\mu =
\dot{x}^\nu \nabla_\nu \dot{x}^\mu = 
\ddot{x}^\mu + \Gamma^\mu_{\nu \rho} \dot{x}^\nu \dot{x}^\rho
\]
of the tangent vectors $\dot{x}^\mu(r)$ along the curve:
\begin{eqnarray*}
(\nabla_{\dot{x}}\dot{x})^t     &=& \Gamma^t_{rt}\dot{x}^r\dot{x}^t+\Gamma^t_{tr}\dot{x}^t\dot{x}^r =2\cdot\frac{1}{2}h^{-1}\dot{h}(-1)(1)=-h^{-1}\dot{h}=h^{-1}\dot{h}\dot{x}^t,  \\
  (\nabla_{\dot{x}}\dot{x})^r &=
&\Gamma^r_{tt}\dot{x}^t\dot{x}^t+\Gamma^r_{rr}\dot{x}^r\dot{x}^r+\Gamma^r_{xx}\dot{x}^x\dot{x}^x+\Gamma^x_{yy}\dot{x}^y\dot{x}^y =\Gamma^r_{tt}+\Gamma^r_{rr}=h^{-1}\dot{h}\dot{x}^r \;.
  \end{eqnarray*}
So all in all we have
\begin{equation}
    \ddot{x}^{\mu}+\Gamma^{\mu}_{\nu\rho}\dot{x}^{\nu}\dot{x}^{\rho}=F(r)\dot{x}^{\mu} \;,
\end{equation}
with $F(r)=h^{-1}\dot{h}$. This shows that the curves
$x^{\mu}(r)$ are geodesics, but the curve parameter $r$
is not an affine parameter. As usual, we can obtain an affine parameter $\lambda$ by insisting on 
a parametrization where the right hand side of the geodesic equation becomes zero. This leads to the relation
\begin{equation}
    \frac{d\lambda}{dr}=e^{\int^r F(s)ds}=h(r) \;,
\end{equation}
which can be integrated to obtain the affine parameter.
For an ingoing transverse null geodesic starting at $t=0, r=R$, the point
$r=0$ is reached at affine distance 
\begin{eqnarray}
    D_A(0,R) &=& \int_R^0 h(r) dr =\int_R^0 \frac{dr}{-2M/r+e^2/r^2} \\ & = &\left[-\frac{2Mr(Mr+e^2)+e^4\ln(|e^2-2Mr|)}{8M^3}\right]_R^0 \;. \label{distance}
\end{eqnarray}
The point $r=0$ has finite affine distance:
\[
0 < D_A(0,R) < \infty \;.
\]
Thus this point is at
`finite distance' for both spacelike and null curves. 

Next we observe that if we take the limit $R\rightarrow r_*$ in \eqref{distance}
we obtain a divergent result due to the logarithmic term. 
This shows, without need for an additional computation that the affine distance 
from any point at $r=R$,  with  $0<R<r_*$ is infinite: 
\[
D_A(R,r_*) = \infty \;.
\]
Moreover, since the solution takes the same form for $r>r_*$ we will find the same divergence  when computing the affine distance from any $r=R$ with $r>R$ to $r_*$. By the same 
logic, the spacelike distance between any such point and the image of the horizon is finite.

\subsection{Interpretation of the singularities}

We will now attempt to give a physical interpretation of the
behaviour of the T-dual solution at $r=0$ and $r=r_*$.

\subsubsection{The T-dual of the singularity, $r=0$}

Let us first summarize our findings:
\begin{itemize}
\item 
The string frame metric degenerates (all metric coefficients go to zero), while
the Einstein frame metric is analytical at $r=0$ and can
be extended analytically to any negative $r$.
\item 
The string frame Ricci curvature diverges, while in the Einstein frame curvature is finite (as is implied by the analyticity of the metric). The Einstein equations imply that the energy momentum tensor is finite (as can of course be checked directly). 
\item 
The dilaton and the axion approach infinity, which is the asymptotic
boundary of their field space:
\[
\varphi' \xrightarrow[r\rightarrow 0+]{} -\infty \;,\;\;\;
\zeta' \xrightarrow[r\rightarrow 0+]{} \pm \infty \;.
\]
The sign of $\zeta'$ depends on the charge $e$, but positive and negative charge
are physically equivalent. The finiteness of the energy mometum tensor, and of the 
Ricci curvature, results from a cancellation between the contributions of the dilaton and the axion. This is possible since due to the timelike T-duality the axion field has 
a kinetic energy term with a flipped sign.
\item 
The position $r=0$ is at finite spacelike and affine distance from points 
at positions $0<r<r_*$.
\end{itemize}
Since in one conformal frame the metric has a curvature singularity while in the other it has not, one might wonder which metric is the `right one.' 
We take the view that the relevant
question is whether the solution taken as a whole, that is involving all fields,
is regular or not. In fact, one of the implications of T-duality is that we should
see the metric on equal footing with all other massless closed string modes. 
As a whole, the T-dual solution is singular at $r=0$, because the dilaton and
axion diverge. The Einstein frame metric happens to be non-singular, because
the conformal transformation between string and Einstein frame, which is 
given by the dilaton, isolates the singularity. 
From a gravitational physicist's point of view we can still 
say that the conformal structure of spacetime is extendable to arbitrary negative 
$r$, though at the expense of singular behaviour of matter fields. This singular
behaviour is not leading to singular stress-energy, but the run-off of scalars
to infinity signals, taking the viewpoint of effective field theory, nevertheless
the breakdown of our model, and the occurance of `new physics'.

From a string theorist's point of view we can say more about this singularity,
by using an embedding of the EM-UHM models into string theory. 
According to the swampland programme, more specifically the swampland 
distance conjecture \cite{Ooguri:2006in}, whenever a scalar field (`modulus') approaches the
asymptotic boundary of its field space (`moduli space'), we should expect that an 
infinite tower of states becomes massless. We will therefore now analyze 
the T-dual solution from this perspective. In the region where $r<r_*$ the
situation does not quite fit the standard assumptions of the swampland programme,
since the scalar manifold of the dilaton-axion system obtained by timelike T-duality
has indefinite signature. But we can still use its metric
to compute the length of the parametrized curve $(\varphi'(r), \zeta'(r))$ which 
corresponds to the T-dual solution. The metric on the scalar manifold is
\[
ds^2 = 2 d \varphi'^2 - e^{2\varphi'} d\zeta'^2 =
2 \frac{d\phi^2}{\phi^2} - \phi^2 d\zeta'^2 =
2 \left( \frac{d\psi^2 - da^2}{\psi^2 } \right) \;,
\]
where we have set $\phi= e^{\varphi'}$,$\psi = 1/\phi$ and $a=\zeta'/\sqrt{2}$
to make explicit that this is the (indefinite) Riemannian metric
on the symmetric space 
$\mathrm{SL}(2,\mathbb{R})/\mathrm{SO}(1,1) \cong \mathrm{AdS}_2$.
For $r\rightarrow 0$ the pull back of this line element along the curve
$(\varphi'(r), \zeta'(r))$ takes the form $ds \propto \frac{dr}{r}$, (up to terms of order $r^0$), which shows that
the length of the curve diverges logarithmically in this limit. Thus the 
field values attained (asymptotically) at $r=0$ are at infinite distance.\footnote{We 
remark that is not unexpected that the curve $(\varphi'(r), \zeta'(r))$ has turned
out to be non-null and thus to have a well defined length, because the solution 
we have T-dualized is non-extremal \cite{Errington:2014bta}. } 
Next, we observe that with our string embedding
the field $\varphi'$ is the
four-dimensional type-IIB$^*$ dilaton, which determines
the four-dimensional string coupling: 
\[
g_{S,(4)} \propto  e^{\varphi'} \;.
\]
Here and in the following we ignore irrelevant constant factors.
Therefore, the singular behaviour for $r\rightarrow 0$
corresponds to vanishing string coupling:
\[ 
\varphi' \xrightarrow[r\rightarrow 0+]{} - \infty \;\;
\Longrightarrow \;\;
g_{S,(4)} \xrightarrow[r \rightarrow 0+]{} 0  \;.
\]
The relation between the four-dimensional and ten-dimensional 
string coupling is
\[
e^{-2 \varphi'} \propto e^{-2 \varphi'_{(10)}} \mathcal{V} \;,
\]
where the volume $\mathcal{V}$ of the internal compact space
depends on various moduli field. With our embedding of the 
dilaton-axion system as the universal part of a Calabi-Yau compactification (or torus), 
all geometric moduli have been consistently 
frozen to constant values. Threrefore the ten-dimenional dilaton
and string coupling show the same behaviour as their four-dimensional
counterparts
\[
\varphi_{(10)} \xrightarrow[r\rightarrow 0+]{} - \infty \;\;
\Longrightarrow \;\;
g_{S,(10)} \propto e^{\varphi_{(10)}} \xrightarrow[r \rightarrow 0+]{} 0  \;.
\]

The 
ten-dimensional gravitational coupling $\kappa$, the Regge slope parameter
$\alpha'$, the string tension $T$ and the ten-dimensional dilaton $\varphi'_{(10)}$ are related by 
\[
\kappa \propto (\alpha')^2 e^{\varphi'_{(10)}} \;,\;\;\;\alpha' = \frac{1}{2\pi T} \;.
\]
Thus $\varphi'_{(10)} \rightarrow -\infty$ at fixed $\kappa$ implies 
that $\alpha' \rightarrow \infty$ and $T\rightarrow 0$. In this limit
the string tension vanishes, and a look at the string mass formula
\[
M^2 \propto T(N_L + N_R + \cdots ) \;,
\]
where $N_{L/R}$ are contributions of left- and right-moving excitations,\footnote{The
ellipses represent further contributions, like normal ordering constants or
contributions from winding modes if we make some dimensions compact.} 
shows that in this limit an infinite tower of states becomes massless. 
This is consistent with the emergent string conjecture \cite{Lee:2019wij}, 
which predicts a collapsing towers of massive states 
resulting from either Kaluza-Klein modes (decompatification 
due to run-away of geometric moduli) or tensionless strings (run-away 
behaviour of the dilaton).

We conclude that the singularity of the T-dual solution is related
to the breakdown of the effective field theory, and in fact of
perturbative string theory, due to strings becoming tensionless. 
This is a new phase of string theory with 
infinitely many massless exitations and enlarged unbroken gauge symmetries. 
One description which has been proposed for this phase are higher-spin field theories, \cite{Sagnotti:2003qa},\cite{Vasiliev:2003ev}, see \cite{Sagnotti:2011jdy} for a more recent review.
It would be interesting to investigate whether these theories can shed
light on what happens at the singularity, in particular whether it can be 
resolved. In our context this question is even more pressing than in 
the standard swampland context, because the singularity, while at infinite
distance in moduli space, is at finite distance in space, and is reached
by light rays at finite affine parameter. It is also tempting to speculate
that understanding the singularity of the T-dual solution may tell us 
something about the singularity of the original solution. Having an equivalence
of string theories (rather than a solution generating technique) requires one
to keep the compactification circle finite, and to consider the effects
of momentum and winding modes. For timelike circles this raises 
additional issues, see \cite{Dijkgraaf:2016lym}, \cite{Blumenhagen:2020xpq}.
It will
be interesting to see whether this is a tractable problem.

\subsubsection{The T-dual of the horizon, $r=r_*$}

We now turn to our main point of interest, the T-dual image of the 
horizon at $r=r_*$. Let us summarize the previous findings
\begin{itemize}
    \item 
    The components $g^S_{tt}$ and $g^S_{rr}$ of the string frame 
    metric diverge, while
    $g^S_{xx}$ and $g^S_{yy}$ are finite. In the Einstein frame, 
    $g^E_{tt}$ and $g^E_{rr}$ have a discontinuity (sign flip at finite value), 
    whereas $g^E_{xx}$ and $g^E_{yy}$ have a zero (thus preserving the signature). 
    \item 
    The Ricci curvature diverges in both frames. As a consequence of the Einstein equations, the stress energy tensor diverges.
    \item 
    The dilaton runs off to infinity
    \[
    \varphi' \xrightarrow[r\rightarrow r_*\mp ]{} + \infty
    \]
    while the axion $\zeta'$ is continuous and in fact analytic at $r=r_*$. 
    \item 
    In the Einstein frame, $r=r_*$ is at finite spacelike and affine distance from any finite points  $r\not= 0, r_*$. In the string frame the distance is finite for spacelike geodesics but infinite for light rays. 
\end{itemize}
At face value, the situation looks more singular than at $r=0$, since the 
metric and curvature are singular in both conformal frames. On the other hand,
we know that the original solution that we have T-dualized is completely
regular at this point. This makes the behaviour at $r=r_*$  a testing ground for whether T-duality with respect to a null direction is meaningful, and whether dynamical transitions between standard and exotic type-II string theories are possible.  The running off of the dilaton, which takes the
same form from both sides, $\varphi' \xrightarrow[r\rightarrow r_*\pm]{} \infty$
is consistent with the idea that the scalar fields have to transfer between
the associated scalar manifolds $SL(2,\mathbb{R})/SO(1,1)$ for $r<r_*$ 
and $SL(2,\mathbb{R})/SO(2)$ for $r>r_*$. The axion $\zeta'$ whose
kinetic term gets flipped in this process is analytic at $r=r_*$, and 
it is manifest that the dilaton travels an infinite distance on its
scalar manifold on either side of the interface.

When repeating the analysis we did for $r=0$, we find that this time 
the four-dimensional string coupling becomes infinite:
\[
\varphi' \xrightarrow[r\rightarrow r_* \pm]{} + \infty \;\;\;
\Longrightarrow \;\;\;
g_{S,(4)}  \propto e^{\varphi'} \xrightarrow[r\rightarrow r_* \pm]{} + \infty \;.
\]
At this point we may invoke S-duality to re-interpret this as a zero coupling
limit of an S-dual theory, leading again to tensionless strings. For four-dimensional
models with $\mathcal{N}=2$ supersymmetry we don't have, in general a (four-dimensional)
S-duality and thus no way to control strong coupling behaviour from a purely
four-dimensional perspective \cite{Medevielle:2021wyx}. However, since due to our embedding the volume
of the internal space is constant, the ten-dimensional dilaton and string
coupling grow as well. We can therefore interprete what happens at $r=r_*$
using ten-dimensional S-duality.
Alternatively, we could choose a toroidal embedding, or choose one of the non-generic
$\mathcal{N}=2$ compactifications which exhibit S-duality. 
For $r>r_*$, the T-dual solution is embedded into type-IIB string theory which
is self-dual under S-duality. S-duality acts as $g_{S,(10)} \rightarrow \frac{1}{g_{S,(10)}}$ on the string coupling and
exchanges fundamental strings with D-strings (D1-branes). Thus the singular
behaviour for $r\rightarrow r_*+$ is related to tensionless D-strings. For 
$r<r_*$, the T-dual solution is embedded into type-IIB$^*$. This theory 
is not self-dual under S-duality, but gets mapped to a different exotic string theory, 
dubbed type-IIB'. The couplings are related by $g_S^{IIB^*} = \frac{1}{g^{IIB'}_S}$
and fundamental strings are replaced by E2-branes, which can be viewed as strings
with a Euclidean worldsheet. Thus the singular behaviour for $r\rightarrow r_* -$
is related to tensionless fundamental IIB'-strings, equivalently, to tensionless
type-IIB$^*$ E2 branes. From this point of view, any proposed continuation from 
$r<r_*$ to $r>r_*$ will need to explain how, in a tensionless limit, D1-strings
can somehow morph into E2-branes. We note that according to the 
refined swampland distance conjecture (emergent string conjecture) \cite{Lee:2019wij}
there is no alternative explanation: due to our choice of string embedding
all geometric moduli are frozen in, preventing an 
explanation in terms of a collapsing Kaluza-Klein tower. Therefore  
tensionsless strings (in a suitable duality frame) are the only remaining 
possible behaviour at the asymptotic boundary of moduli space.

\section{Dualizing a black hole solution (from type-IIA* to type-IIB/IIB*)}

In this section we consider another pair of T-dual solutions which 
is closely related to the previous. In \cite{Gutowski:2020fzb}
it was observed that the EM$_-$ theory, that is Einstein-Maxwell
theory with a flipped sign for the Maxwell term, admits a planar
black hole solution which is related to the planar cosmological
solution of the standard EM$_+$ theory by analytic continuation 
in the charge parameter $e$. 
According to \cite{Medevielle:2021wyx}, we can embedd this solution 
into the twisted EM-UHM model, and further into
toroidal or Calabi-Yau compactifications of type-II$^*$ string theories,
see Figure \ref{fig:2}.  For definiteness we assume an embedding into type-IIA$^*$. Since scalars, 
including the dilaton, are trivial, string frame and Einstein frame coincide. 
The non-trivial fields are the metric and Maxwell field, given by
\begin{equation}
\label{Planar_BH}
\begin{array}{c}
d s^{2}=+f(r) d t^{2}-\frac{d r^{2}}{f(r)}+r^{2}\left(d x^{2}+d y^{2}\right), \quad f(r)=\left(-\frac{2 M}{r}+\frac{e^{2}}{r^{2}}\right) \;, \\
 F=-\frac{e}{r^{2}} d t \wedge dr \;.
\end{array}
\end{equation}
As discussed in detail in \cite{Gutowski:2020fzb} the maximal analytic extension
of this metric has the same conformal structure as the maximal analytic extension of the 
Schwarzschild metric. 
The geometry is of course different: the spatial symmetry is planar rather than spherical, and the solution is not asymptotically flat, but asymptotic to the static 
version of the type-AIII solution of Einstein gravity,
which describes a vacuum outside an infinitely extended 
homogeneous planar mass distribution. The non-static version of
the type-AIII solution is the Kasner cosmological solution
which desribes the timelike asymptotics of the planar
Einstein-Maxwell solution see \cite{Griffiths_Podolsky} and also \cite{Aniceto:2019rhg}
for a detailed description of these solutions.

For $r>r_*$ the Killing vector field $\xi=\partial_t$ is timelike, so that the exterior
region is static. For  $r<r_*$ the Killing vector field becomes spacelike, and all
future pointing timelike and null geodesic reach the spacelike singularity $r=0$
at finite affine parameter. To be precise, $-\infty < t,x,y <\infty$, $0<r<\infty$
covers half of the maximally extended solution, and we have chosen the time direction 
such that this part corresponds to a black hole rather than a white hole. The remaining half of the maximally extended spacetime
is a time reversed version of the black hole, that is, a white hole.

We now T-dualize the black hole solution with respect to $\xi=\partial_t$.
In the internal, non-static patch $0<r<\infty$, the T-duality is spacelike. 
The dual theory is therefore again the twisted gravity-dilaton-axion system, which embeds into the twisted 
EM-UHM model, and further 
into type-IIB$^*$ string theory. The  dual string frame metric is
\begin{align}
    ds'^2_{Str}&=h(r) \left(d t^{2}- d r^{2}\right) +r^{2}\left(d x^{2}+d y^{2}\right) \nonumber  \\
    &=\frac{dt^2 - dr^2}{\left(-\frac{2 M}{r}+\frac{e^{2}}{r^{2}}\right)} +r^{2}\left(d x^{2}+d y^{2}\right) \;,
\end{align}
and the R-R sector is the same as for the dualized cosmological solution discussed in the previous section.
In the Einstein frame the metric takes the form:
\begin{equation}
    ds'^2=dt^2-dr^2+(e^2-2Mr)(dx^2+dy^2) \;.
\end{equation}
As previously one can check explicitly that this is a solution to the equations of motion. 
The dilaton has the same profile as before.  
In the Einstein frame we have 
\begin{equation}
    R^{(E)}_{\mu\nu}=\frac{2M^2}{(e^2-2Mr)^2}\delta_{rr} \;.
\end{equation}
The Einstein frame Ricci scalar is 
\begin{equation}
    R^{(E)} =-\frac{2M^2}{(e^2-2Mr)^2} \;,
\end{equation}
and the string frame Ricci tensor and scalar are
\begin{eqnarray}
    R^{(S)}_{\mu\nu}&=&\text{diag}\Big(\frac{e^4-2e^2Mr+2M^2r^2}{r^2(e^2-2Mr)^2},\frac{3e^4-10e^2Mr+6M^2r^2}{r^2(e^2-2Mr)^2}, \nonumber  \\
    && -\frac{e^2-2Mr}{r^2},-\frac{e^2-2Mr}{r^2}\Big)\;, \\
    R^{(S)}&=&\frac{4M^2}{r^2(e^2-2Mr)} \;.
\end{eqnarray}
The equations for spacelike and affine distances remain 
essentially unchanged, and one finds the same 
behaviour as in the dualized cosmological solution. 
Therefore singularities at $r=0$ and $r=r_*$ are again
related to the presence of tensionless strings. 
We note that the locus $r=0$ is T-dual to a spacelike rather than 
timelike singularity, a type of singularity that is poorly
understood in string theory.

\section{Relating cosmologies to black holes by T-duality (from type-IIA to type-IIA$^*$)?}

We now turn our attention to
the relation between the cosmological and the black hole solution. Type-IIA and type-IIA$^*$ string theory are related by the compostion of 
two T-dualities, one timelike, the other spacelike. The effect 
on the Einstein-Maxwell subsector in the four-dimensional effective field theory 
is a sign flip of the 
Maxwell term. As discussed in \cite{Gutowski:2020fzb} one can
write static, planar symmetric solutions of the EM and twisted EM-theory
uniformly as 
\[
ds^2 = - f(r) dt^2 + f(r)^{-1} dr^2 + r^2 (dx^2 + dy^2) \;,\;\;
f(r) = \frac{2c}{r} + \frac{\epsilon e^2}{r^2} \;,
\]
with $\epsilon=-1$ for EM$_+$ and $\epsilon=1$ for EM$_-$.
The parameter $c$
is an integration constant, which a priori can be positive or negative. 
For fixed $c$ the solutions are related by analytic continuation of the
charge $e \rightarrow ie$, reflecting that the sign flip of the 
Maxwell term is equivalent to making gauge fields imaginary. 

Imposing the absence of naked singularities requires that $f(r)$ must have a zero at some point $r_*>0$,
which implies that 
\[
f(r) = \epsilon \left( \frac{2M}{r} - \frac{e^2}{r^2}\right) \;,
\]
where $M>0$ and $M^2>e^2$. 
Thus the type-IIA and type-IIA$^*$ solutions with horizons are related by
$f(r) \rightarrow - f(r)$.
We observe that when restricting the parameters $M,e$, in order to 
guarantee the existence of a horizon, flipping the 
sign $\epsilon$ is no longer equivalent to the
analytic continuation $e \rightarrow ie$ of the charge, but
requires in addition to flip the sign of the mass $M\rightarrow-M$.
Equivalently, we could flip the sign of the transverse 
coordinate $r\rightarrow -r$. Viewed this way, we can 
formally join the cosmological and black hole solution 
at the singularity $r=0$. This raises the question 
whether we could use a chain of dualities and analytic
continuations to glue both solutions 
together into a single spacetime.
Such a construction would be similar in spirit to the proposal
to put a cosmological solution inside a black hole, using 
an S-brane as interface \cite{Brandenberger:2021jqs}. In our case
the interface would be provided by a type-II$^*$ E-brane instead of 
an S-brane.

The obvious candidate for such a chain of dualities is combining 
a spacelike with a timelike T-duality, the minimum
required to get from type-IIA to type-IIA$^*$. While
this relates the two actions to one another, the solutions
themselves only have one (relevant) isometry, which is
spacelike in some regions and timelike in others. 
This leaves one to apply spacelike T-duality 
in a static region, continue through the horizon, and 
then apply timelike T-duality. Equivalently, we can
use that we have already applied T-duality to both 
solutions in both regions, and compare the resulting
type-IIB/IIB$^*$ solutions with one another. The 
string frame metrics that we have obtained are:
\begin{eqnarray*}
    ds^{2,(S)}_{IIA \rightarrow IIB^*/IIB} &=& 
    \frac{-dt^2 + dr^2}{f(r)} + r^2 (dx^2 + dy^2) \;, \\
    ds^{2,(S)}_{IIA^* \rightarrow IIB/IIB^*} &=& \frac{dt^2 - dr^2}{f(r)} + r^2 (dx^2 + dy^2) \;.
\end{eqnarray*}
where $r>0$ and 
$f(r) = - \frac{2M}{r} + \frac{e^2}{r^2}$ with $M>0$ and $M^2>e^2$.
As real solutions, these are two distinct families that cannot
be combined into a single spacetime (even when allowing to patch
solutions formally across singularities). Like for the 
corresponding IIA/IIA$^*$ solutions, gluing along $r=0$ requires an
additional analytic continuation in the charge. Therefore, 
the solutions are not connected by a combination of T-duality 
and continuation in spacetime.

There is one further option to be explored in the future, 
namely that both IIB/IIB$^*$ solutions arise as special members
of a larger family of solutions (with generic members having
less isometries). To motivate this idea,
recall that when saying that D$p$ branes `are' 
T-dual to D$(p-1)$ branes, 
one actually refers to a process where one first T-dualizes a D$p$-brane along
an isometric direction, and then allows that the dual solutions depends explicitly
on the coordinate of that direction, thus obtaining
a localized D$(p-1)$ brane solution with a reduced number of isometries. Conversely, going back 
from a localized D$(p-1)$ brane to a D$p$-brane involves as a first step to create
an isometry by de-localization, resulting in a smeared D$(p-1)$ which can then be T-dualized into
a D$p$-brane.\footnote{See for example 
\cite{Mohaupt:2000gc} for a pedagogical review.}
We should therefore investigate whether 
the type-IIB/IIB$^*$ solutions obtained by T-dualizing type-IIA/IIA$^*$ solutions 
admit localized versions, and whether one can move from IIA to IIA$^*$ by a chain
of T-dualities combined with localizations/delocalizations and extensions in spacetime.

While it is not obvious how to obtain localized versions of the type-IIB/IIB$^*$ solutions,
we can give an argument why they should exist:
one of the dimensional uplifts of the IIA solution 
is the D4-D4-D4-D0 system. It was shown in \cite{Behrndt:1997ch} that this brane configuration can be T-dualized, after
Wick rotation, into the D3-D3-D3-D(-1) brane system of type-IIB. This involves 
a localization of precisely the type mentioned before. Adapting the procedure of
\cite{Behrndt:1997ch} requires two modifications: (i) applying timelike T-duality 
instead of a combination of Wick rotation and spacelike T-duality, which is 
straightforward, thus obtaining a IIB$^*$ E-brane system, which can be further
T-dualized into a IIA$^*$ E-brane system, (ii) generalizing the localization/delocalization
procedure to non-extremal solutions. As mentioned before, the IIA/IIA$^*$ solutions only 
have horizons as long as they are non-extremal, $M>|e|>0$. The creation of an isometry 
by smearing can be viewed of the continuous limit of a periodic array of localized 
solutions, but this makes use of the no-force property of BPS (hence, extremal) solutions.
One way to adapt the construction to non-extremal solutions is to first take the extremal 
limit, perform (de-)localization, and then to make the solution non-extremal again. We leave working 
out the details to future work.

\section{Conclusions and Outlook}

In this paper we have derived four-dimensional Buscher rules for spacelike and timelike T-duality for Einstein-Maxwell theory coupled to the universal hypermultiplet 
as well as for the twisted version of this theory. We have 
applied these results to study the effect of T-duality 
on solutions which contain a non-extremal Killing horizon as well as a curvature singularity. While we
have considered specific examples for concreteness, our findings result from qualitative properties of the 
function $f(r)$ and therefore are expected to hold for a large class of spacetimes.
Using embeddings
into type-II/II$^*$ string theory, we found that the singularities of the T-dual solutions occur at the boundary of the underlying moduli spaces, and are related to tensionless strings. This opens various venues for the
further study of such singularities, and, possibly, also
the singularities of the original solutions that we have T-dualized.

We think there are three promising directions to explore.
\begin{enumerate}
\item 
Using the embedding into type-II/II$^*$ string theory, 
we can study the higher-dimensional brane configurations
underlying our solutions, as well as others. This is the
most direct way to explore their microscopic, stringy
nature. It will also allow to decide whether type-II cosmological and type-II$^*$ black holes solutions can be related by a chain of dualities. More generally, one could extend this to investigating the global geometric structures 
of non-extremal branes of standard and of exotic type-II theories, as well as the 
action of T-duality on them.

\item 
Since the behaviour of the T-dual solutions at their singularities involves tensionless strings, higher spin
field theories may be helpful to understand the behaviour
in the phase the solutions are driven into.
\item 
Generalized geometry, as well as double field theory 
are formalisms which allow for a fully T-duality 
covariant description, without the need to take the 
full string spectrum into account. The simplest option, and
the one which is arguably the best understood mathematically, is generalized geometry, which is based on 
an extension of the tangent bundle, while leaving spacetime
as is. In this formalism one works with a finite 
compactification circle, which may allow one to draw 
conclusions on a solution using its T-dual. It will
be interesting to explore whether some of the singular
behaviour seen in a Riemannian geometry set-up gets 
mitigated when working with generalized geometry data.
\end{enumerate}
Work along these lines will be able to shed light on the status of timelike
T-duality and of type-II$^*$ string theories, as well as type-II
string theories in non-Lorentzian signatures. While we have been
able to argue that the singular behaviour of the T-dual solution 
at the image of the horizon is due to the presence of tensionless strings,
it is not clear whether the solution can be continued through the
`tensionless string interface', which would amount to a dynamical transition from 
type-IIB to type-IIB$^*$ string theory. One option would be that the 
type-IIB moduli space ends in a tensionless string phase, without 
an extension into the exotic type-IIB$^*$ branch. This would support the case that
type-IIB$^*$ theories, while they can be generated by formally applying
T-duality, are not dynamically connected to standard string theory. 
Conversely, establishing a dynamical connection in a single example
would force one to address the various exotic features, potential benefits,
but also potential pathologies of the exotic type-II theories 
\cite{Dijkgraaf:2016lym},\cite{Blumenhagen:2020xpq}.

The issue of dynamical type-II/type-II$^*$ transitions is different from, 
but closely related to the issue of dynamical signature change and thus
transitions between type-II theories in different signatures. We noted before that, by means of dimensional reduction with respect to the Killing vector
field, any non-extermal Killing horizon provides an example of `smooth 
signature change': the dimensionally reduced effective field theories
have Euclidean and Lorentzian signature, respectively, and while the
transition between them cannot be smooth, there is a perfectly well
behaved dimensional lift where the singular interface becomes a
horizon. Similarly, \cite{Dijkgraaf:2016lym} have argued that 
dynamical signature change between ten-dimensional type-II superstrings
is possible, by relating them through eleven dimensional M-theory (which 
also has two exotic versions with non-Lorentzian signature \cite{Hull:1998ym}).
Thus the same methodology can be applied for transitions between
type-II/type-II$^*$ and for signature change. Establishing or refuting
the viability of either type of dynamical transition is important
for understanding `how big' the configuration space and symmetry 
group underlying a non-perturbative formulation of string theory 
needs to be. Some approaches to uncovering a unifying symmetry
principle, such as compactification of all direction including 
time \cite{Moore:1993zc}, 
exceptional field theory
and the $E_{11}$-approach \cite{Keurentjes:2004bv}, \cite{Hohm:2019bba}, \cite{Tumanov:2015iea}, 
naturally include timelike reductions and exotic type-II theories. 

While we
have given a tentative interpretation of the T-dualised solutions from the
perspective of the emergent string conjecture \cite{Lee:2019wij}, it may 
also be interesting to analyse them using the swampland cobordism conjecture
\cite{McNamara:2019rup}. Situations where singularities appear
at finite distance in spacetime but infinite distance in moduli space have been
interpreted as examples of dynamical cobordism where spacetime ends due
to stringy effects (such as tachyon condensation) in an `end of the world brane'
or `wall of nothing' \cite{Buratti:2021yia,Buratti:2021fiv,Angius:2022mgh}. 
It should be interesting to investigate whether the solutions
found in this paper fit into this framework, and whether they correspond to
ends of spacetime (`cobordisms to nothing') or transitions between spacetimes 
(`interpolating cobordisms').\footnote{This line of thought was suggested to us
by Andriana Makridou.} 

Finally, it is interesting to note that in the stretched horizon approach to 
black hole entropy, a gas of high temperature, and therefore effectively 
tensionless strings play a central role \cite{Susskind:1993if}. This raises the question whether
the tensionless strings associated to T-dual horizons describe the same
physics in a different duality frame. Which in turn brings us to the
question whether what we have observed about the solutions obtained and
studied in this paper has direct relevance for the original solutions, and, in
particular, on the physics of horizons. Since
we have used T-duality as a solution generating technique, the results obtained
in this paper do not allow us to make any such claims, but they could be used as a basis for
further investigations. To obtain an equivalence between string backgrounds, one
would need to work with a finite-size circle and thus one would need to control 
the associated momentum and winding notes. Moreover,
in the case of timelike T-duality, this circle would introduce closed timelike curves,
which raises the question whether to simply discard such solutions as unphysical. Since
string theory aims to be a fundamental theory, it does not seem satisfactory to exclude
 such solutions by postulating an ad hoc `chronology protection principle,' but rather 
 the full dynamics should imply such a principle. An example
 of such a mechanism was given in \cite{Gimon:2004if} 
 and dubbed `holographic chronology
 protection.' Interestingly this example is based on a construction involving
 D1- and D5-branes and thus related to one of the ten-dimensional lifts of the
 solutions we have T-dualized.


\subsection*{Acknowledgements}

We thank Vicente Cort\'es, Chris Hull, Andriana Makridou and L\'arus Thorlacius for 
illuminating discussions. We also thank the referee for their constructive
feedback on the first version of this paper, and for 
bringing reference \cite{Gimon:2004if} to our attention. 
T.M. would like to thank
the Isaac Newton Institute for Mathematical Sciences
for support and hospitality during the programme `Black
Holes: a bridge between number theory and holographic
quantum information' when work for this paper was
undertaken. This work was supported by EPSRC grant
number EP/R014604/1. T.M. would like to thank the
Department of Mathematics of the University of Hamburg
for support and hospitality during several visits 
when work for
this paper was undertaken. The work of M.M. is supported by an EPSRC DTP International
Doctoral Scholarship (project number 2271092) and JSPS Postdoctoral Fellowships for Research in Japan (Standard) No. P23774.
The work of T.M. was supported by the STFC Consolidated Grant ST/T000988/1.
{\em Data access statement:} There is no additional data beyond what is included
in this paper. 

\appendix

\section{Details of selected computations}

\subsection{Details for the Hodge duality between universal axion and
Kalb-Ramond field}

We start from the action (\ref{action3a}) and introduce a two-form potential
\[
\hat{B} = \frac{1}{2!} \hat{B}_{\mu \nu} dx^\mu \wedge dx^\nu \;,\;\;
\]
with associated three-form field strength\footnote{$A_{[\mu_1 \cdots \mu_n]}$ denotes antisymmetrization of indices, that is summing over all permutations, weighted by the signum of the permutation, and 
normalized by applying a factor $(n!)^{-1}$.}
\[  
\hat{H} = \frac{1}{3!} \hat{H}_{\mu \nu \rho} dx^\mu \wedge dx^\nu  \wedge dx^\rho
= d B  \Rightarrow H_{\mu\nu\rho} = 3 \partial_{[\mu } B_{\nu \rho]} \;.
\]
In order to replace the gradient $\partial_\mu \tilde{\varphi}$ by a general one-form $\hat{F}=\hat{F}_\mu dx^\mu$, we use the two-form $\hat{B}$
subject to  the Bianchi identity 
$d\hat{F}=0 \Leftrightarrow \partial_{[\mu} \hat{F}_{\nu]}$ which guarantees the
(local) existence of a potential $\tilde{\varphi}$. This is done
by adding the term
\[
S_{\mathrm{mult}} = - (3!)K \int \hat{B} \wedge d\hat{F} = (3!)K \int \hat{H} \wedge \hat{F}
\]
to the action, where $K$ is a numerical constant that we can set
to a convenient value later. In terms of components:
\[
S_{\mathrm{mult}} = (3!) K \int \frac{1}{3!} \hat{H}_{\mu \nu \rho} 
\hat{F}_\lambda \; dx^\mu \wedge dx^\mu \wedge dx^\rho \wedge dx^\lambda \;.
\]
Note that as an integral of a four-form over space-time, this is
defined independently of the metric. We now re-express this as an
integral of a function against the volume element assocated with 
the string frame metric:
\begin{eqnarray*}
S_{\mathrm{mult}} &=&  K \int  \hat{H}_{\mu \nu \rho} 
\hat{F}_\lambda \, \delta^{\mu \nu \rho \sigma}_{0123} \,
dx^0 \wedge dx^1 \wedge dx^2 \wedge dx^3  \\
&=& K \int d^4x \sqrt{\hat{g}_S}  \; \epsilon^{\mu \nu \rho \sigma}
 \hat{H}_{\mu \nu \rho} 
\hat{F}_\lambda \;.
\end{eqnarray*}
Here 
\[
\delta^{\mu \nu \rho \sigma}_{0123} = 4!
\delta^{[\mu}_0 \delta^\mu_1 \delta^{\rho}_2 \delta^{\lambda]}_3
= \left\{ 
\begin{array}{ll} 
1, & \mathrm{if\;} (\mu, \nu, \rho, \sigma)\; \mathrm{is\;an\;even\; permutation \; of \;} (0,1,2,3)\;,  \\
- 1, &\mathrm{if\;} (\mu, \nu, \rho, \sigma)\; \mathrm{is \; an \; \; odd \; permutation \; of \;} (0,1,2,3) \;, \\
0, & \mathrm{else} \;,
\end{array} \right.
\]
is the numerical permutation symbol (which transforms as a 
tensor density),
while
\[
\epsilon^{\mu \nu \rho \sigma} = \frac{1}{\sqrt{\hat{g}_S}} 
\delta^{\mu \nu \rho \sigma}_{0123}
\]
is the associated tensor.

Now we replace \eqref{action3} by
\begin{eqnarray}
S[\hat{F}, \hat{B}] &=& 
\int d^4x\sqrt{\hat{g}_S}(-2)e^{2\varphi}\left[\right. \hat{F}_\mu
+\frac{1}{2}(\zeta\partial^{\mu}\tilde{\zeta}-\tilde{\zeta}\partial^{\mu}\zeta)]\left[\right. \hat{F}^\mu +\frac{1}{2}(\zeta\partial_{\mu}\tilde{\zeta}-\tilde{\zeta}\partial_{\mu}\zeta)] \label{action4}  \nonumber  \\
&& 
+K \int d^4 \sqrt{\hat{g}_S} \, \epsilon^{\mu\nu\rho\lambda}\hat{H}_{\mu\nu\rho} \hat{F}_\lambda 
\;.
\end{eqnarray}
Variation with respect to $\hat{B}_{\mu \nu}$
imposes the Bianchi identity $\partial_{[\mu} 
\hat{F}_{\nu]}=0$, which is solved by $\hat{F}_\nu=
\partial_\nu \tilde{\varphi}$. Upon substituting
this into \eqref{action4} we recover \eqref{action3a}. To obtain the dualized action, we instead vary with respect to $\hat{F}_\mu$ and obtain
\begin{equation}
        \hat{F}^{\sigma}=-\frac{1}{2}(\zeta\partial^{\sigma}\tilde{\zeta}-\tilde{\zeta}\partial^{\sigma}\zeta)+\frac{1}{4} Ke^{-2\varphi}\epsilon^{\mu\nu\rho\sigma}\hat{H}_{\mu\nu\rho} \;.
\end{equation}
Plugging this back into the action, upon
choosing $K=\frac{1}{3}$, we finally arrive at  the dual action (\ref{action5}).

\subsection{Details for the T-duality transformation of $\partial_\mu \tilde{\zeta}$}

\begin{align}
    \partial_{\mu}\tilde{\zeta}&\rightarrow \partial_{\mu}\tilde{\xi} \nonumber \\
    &\Rightarrow \underbrace{\epsilon_{\mu\nu\lambda}\epsilon^{\mu\nu\rho}}_{=-2\delta^{\rho}_{\lambda}}\partial_{\rho}\tilde{\xi}=-\epsilon_{\mu\nu\rho}e^{\sigma}(F^{*\mu\nu}+\xi V^{\mu\nu}) \nonumber\\
    &\Rightarrow \partial_{\lambda}\tilde{\xi}=\frac{1}{2}\epsilon_{\mu\nu\lambda}e^{\sigma}(F^{*\mu\nu}+\xi V^{\mu\nu}) \nonumber\\
    &=\frac{1}{2}\epsilon_{\mu\nu\lambda}e^{\sigma}(F^{\mu\nu}-2\partial_{[\mu}\xi V_{\nu]}) \nonumber \\
    &=\frac{1}{2}\epsilon_{\mu\nu\lambda}e^{\sigma}\left(\partial^{\mu}A^{\nu}-\partial^{\nu}A^{\mu}-2\partial_{[\mu}\xi V_{\nu]}\right) \nonumber\\
    &=\frac{1}{2}\epsilon_{\mu\nu\lambda}e^{\sigma}\left(\partial^{\mu}(A^{\nu}+\xi V^{\nu})-\partial^{\nu}(A^{\mu}+\xi V^{\mu})-\partial_{\mu}\xi V_{\nu}+\partial_{\nu}\xi V_{\mu}\right)\nonumber\\
    &=\frac{1}{2}\epsilon_{\mu\nu\lambda}e^{\sigma}\left(\partial^{\mu}\hat{A}^{\nu}+\partial^{\mu}(\xi V^{\nu})-\partial^{\nu}\hat{A}^{\mu}-\partial^{\nu}(\xi V^{\mu})-\partial_{\mu}\xi V_{\nu}+\partial_{\nu}\xi V_{\mu})\right)\nonumber\\
    &=\frac{1}{2}\epsilon_{\mu\nu\lambda}e^{\sigma}\left(\hat{F}^{\mu\nu}+\xi\partial^{\mu}V^{\nu}-\xi\partial^{\nu}V^{\mu}\right) \nonumber\\
    &=\frac{1}{2}\epsilon_{\mu\nu\lambda}e^{\sigma}\left(\hat{F}^{\mu\nu}+\xi V^{\mu\nu}\right) \nonumber\\
    &=\frac{1}{2}\epsilon_{\mu\nu\lambda}e^{\sigma}\left(\hat{F}^{\mu\nu}+\hat{A}_{y}(\partial^{\mu}(e^{-2\sigma}\hat{g}^{\nu y})-\partial^{\nu}(e^{-2\sigma}\hat{g}^{\mu y}))\right) \nonumber\\
    &=\frac{1}{2}\hat{\epsilon}_{y\mu\nu\lambda}\sqrt{\hat{g}_{yy}}\left[\hat{F}^{\mu\nu}+\hat{A}_{y}\left(\partial^{\mu}\left(\frac{\hat{g}^{\nu y}}{\hat{g}_{yy}}\right)-\partial^{\nu}\left(\frac{\hat{g}^{\mu y}}{\hat{g}_{yy}}\right)\right)\right] \;.
\end{align}

\subsection{Details for the T-duality transformation of the Maxwell field}
\begin{align}
    \hat{F}_{\mu\nu}&=\partial^{\mu}\hat{A}^{\nu}-\partial^{\nu}\hat{A}^{\mu}=\partial^{\mu}(A^{\nu}-\xi V^{\nu})-\partial^{\nu}(A^{\mu}-\xi V^{\mu}) \nonumber\\
    &=F^{*\mu\nu}=-e^{-\sigma}\epsilon^{\mu\nu\rho}\partial_{\rho}\tilde{\xi}-\xi V^{\mu\nu} \nonumber\\
    &\rightarrow-e^{\sigma}\epsilon^{\mu\nu\rho}\partial_{\rho}\tilde{\zeta}-\zeta(-F'^{\mu\nu}) \nonumber\\
    &=-e^{\sigma}\epsilon^{\mu\nu\rho}\partial_{\rho}\tilde{\zeta}+\zeta\left(\partial^{\mu}A'^{\nu}-\partial^{\nu}A'^{\mu}\right) \nonumber\\
    &=-e^{\sigma}\epsilon^{\mu\nu\rho}\partial_{\rho}\tilde{\zeta}+\zeta\left(\partial^{\mu}(\hat{B}_{\nu y})-\partial^{\nu}(\hat{B}_{\mu y})\right) \nonumber\\
    &=-\sqrt{\hat{g}_{yy}}\hat{\epsilon}^{y\mu\nu\rho}\partial_{\rho}\tilde{\zeta}+\zeta\left(\partial^{\mu}(\hat{B}_{\nu y})-\partial^{\nu}(\hat{B}_{\mu y})\right) \;.
\end{align}


\providecommand{\href}[2]{#2}\begingroup\raggedright\endgroup

\end{document}